\documentclass[english,nofootinbib]{revtex4}
\usepackage[T1]{fontenc}
\usepackage[latin9]{inputenc}
\usepackage{babel}
\usepackage{amsmath}
\usepackage{amssymb}
\usepackage{graphicx}
\usepackage{esint}
\usepackage[unicode=true,pdfusetitle,
 bookmarks=true,bookmarksnumbered=false,bookmarksopen=false,
 breaklinks=false,pdfborder={0 0 1},backref=false,colorlinks=false]
 {hyperref}

\makeatletter
\@ifundefined{textcolor}{}
{%
 \definecolor{BLACK}{gray}{0}
 \definecolor{WHITE}{gray}{1}
 \definecolor{RED}{rgb}{1,0,0}
 \definecolor{GREEN}{rgb}{0,1,0}
 \definecolor{BLUE}{rgb}{0,0,1}
 \definecolor{CYAN}{cmyk}{1,0,0,0}
 \definecolor{MAGENTA}{cmyk}{0,1,0,0}
 \definecolor{YELLOW}{cmyk}{0,0,1,0}
}

\usepackage{mathpazo}
\usepackage{slashed}

\makeatother

\begin{document}

\title{Quadratic Isocurvature Cross-Correlation, Ward Identity, and Dark
Matter}

\author{Daniel\ J.H.\ Chung, Hojin Yoo, and Peng Zhou\\
Department of Physics, University of Wisconsin\\
1150 University Avenue, Madison, WI 53706, USA}
\begin{abstract}
Sources of isocurvature perturbations and large non-Gaussianities
include field degrees of freedom whose vacuum expectation values are
smaller than the expansion rate of inflation. The inhomogeneities
in the energy density of such fields are quadratic in the fields to
leading order in the inhomogeneity expansion. Although it is often
assumed that such isocurvature perturbations and inflaton-driven curvature
perturbations are uncorrelated, this is not obvious from a direct
computational point of view due to the form of the minimal gravitational
interactions. We thus compute the irreducible gravitational contributions
to the quadratic isocurvature-curvature cross-correlation. We find
a small but non-decaying cross-correlation, which in principle serves
as a measurable prediction of this large class of isocurvature perturbations.
We apply our cross-correlation result to two dark matter isocurvature
perturbation scenarios: QCD axions and WIMPZILLAs. On the technical
side, we utilize a gravitational Ward identity in a novel manner to
demonstrate the gauge invariance of the computation. Furthermore,
the detailed computation is interpreted in terms of a soft-$\zeta$
theorem and a gravitational Ward identity. Finally, we also identify
explicitly all the counterterms that are necessary for renormalizing
the isocurvature perturbation composite operator in inflationary cosmological
backgrounds.

\tableofcontents{}
\end{abstract}
\maketitle

\section{Introduction}

As physics beyond the Standard Model is expected to contain many fields
in addition to the inflaton, there are many candidates for isocurvature
perturbations in the context of inflationary cosmology, including
those of the dark matter. Indeed, the current data is consistent with
the existence of an $O(5\%)$ isocurvature component \cite{Crotty:2003fp,Bean:2006qz,Komatsu:2008ex,Valiviita:2009ck,Sollom:2009vd,Komatsu:2010in,Hinshaw:2012fq,Ade:2013rta}.
Furthermore, it is well known that quadratic isocurvature perturbations
(i.e. the vacuum expectation value of the field is much smaller than
the Hubble expansion rate) are one of the very few ways to generate
measurably large local non-Gaussianities \cite{Linde:1996gt,Kofman:1989ed,Bartolo:2001cw,Geyer:2004bx,Ferrer:2004nv,Boubekeur:2005fj,Barbon:2006us,Lyth:2006gd,Koyama:2007if,Lalak:2007vi,Huang:2007hh,Lehners:2008vx,Beltran:2008tc,Kawasaki:2008jy,Langlois:2008vk,Kawasaki:2008sn,Takahashi:2009cx,Langlois:2009jp,Byrnes:2010em,Chen:2010xka,Langlois:2010fe,Chung:2011xd,Langlois:2011zz,Mulryne:2011ni,Gong:2011cd,DeSimone:2012gq,Enqvist:2012vx,Kawasaki:2011pd,Assadullahi:2013ey,Kawasaki:2013ae,Langlois:2013dh,Nurmi:2013xv}
in the context of the slow-roll inflationary paradigm. The only nontrivial
requirement that the isocurvature field degree of freedom must possess
is that it be light enough to be excited by the inflationary quasi-de
Sitter (dS) background and that it not be conformally invariant. In
the literature \cite{Chung:2004nh,Beltran:2006sq,Hertzberg:2008wr},
quadratic isocurvature perturbations are often assumed to have negligible
cross-correlations with the curvature perturbations (which corresponds
to the inflaton field degree of freedom dressed by gravity). However,
the gravitational interactions lead to a minimum cross-correlation,
which in principle can be observationally important. We present a
computation of this minimal gravitational cross-correlation in this
paper. 

As explained below, the form of the gravitational interaction between
the curvature and isocurvature perturbations naively suggests that
there can be cross correlators which do not vanish in the long wavelength
limit. If this was true, the cross correlation can dominate over the
isocurvature two-point function in the observables since the latter
vanishes in the long wavelength limit for a massive field. By an explicit
rigorous computation, we show that the cross correlator vanishes in
the long wavelength in such a way that the cross correlation induced
by gravity never dominates over the isocurvature two-point function,
given that the curvature inhomogeneity perturbation is characterized
by a strength of order $10^{-5}$. We explain this qualitatively as
well using a combination of a soft-$\zeta$ theorem \cite{Maldacena:2002vr,Weinberg:2003sw,Hinterbichler:2012nm,Cheung:2007st,Maldacena:2011nz,Assassi:2012zq,Li:2008gg,Antoniadis:2011ib,Kehagias:2012pd,Creminelli:2012qr,Creminelli:2012ed,Senatore:2012wy,Sugiyama:2012tr,Assassi:2012et,Creminelli:2011mw,Creminelli:2011rh,Kehagias:2013yd,Li:2010yb}
and a Ward identity associated with a spatial dilatation diffeomorphism.
We also check the gauge invariance of our computation using a Ward
identity. 

Among the possible isocurvature candidates, thermal dark matter is
usually produced copiously by the inflaton decay products, which typically
leads to a large suppression of isocurvature effects. On the other
hand, nonthermal dark matter that is not produced by the inflaton
decay can easily generate large isocurvature effects that survive
until today. Hence, as an illustration, we apply our computation of
the cross correlation to two different nonthermal dark matter models:
QCD axions and WIMPZILLAs. In both cases, we find a cross-correlation
characterized by the parameter $\left|\beta\right|\sim O(10^{-5})$
(the parameter definition is given in Eq.~(\ref{eq:fractional_cross-correlation}))
which is below the boundary value of $O(10^{-2})$ when the cross
correlation becomes competitive with the isocurvature two-point function.
In principle, $\beta$ can be measured and is a generic prediction
of this class of nonthermal dark matter quadratic isocurvature models.
Note that even though the nonthermal dark matter fields can be identified
with the isocurvature degrees of freedom, this scenario is consistent
with the WIMP dark matter scenario since the isocurvature perturbations
can be as small as an order $10^{-5}$ fraction of the total dark
matter and still leave an isocurvature imprint on the CMB spectrum. 

The order of presentation is as follows. In Section \ref{sec:Curvature-and-Isocurvature},
we present our assumptions about the inflationary cosmology, review
gauge invariant variables in the perturbation theory, and summarize
the observational constraints on the isocurvature scenario relevant
to our paper. One of the most important aspects of this section is
our review of features of the $\beta$ variable that we compute. In
Section \ref{sec:Computation-of-Correlators}, we first explain two
naive estimates, one leading to the wrong observationally large result,
and the other leading to the correct suppressed result. In explaining
the correct estimate (which requires assumptions that cannot be known
without the justification of a full computation), we present the interpretation
in terms of a soft-$\zeta$ theorem and a Ward identity. The rigorous
explicit computation at one loop is then presented, demonstrating
how the correct naive estimate result is achieved. We also present
in this section how gauge invariance is achieved for these quadratic
isocuvature computations using a gravitational Ward identity. Next,
we apply these results to the axion and the WIMPZILLA scenarios in
Section \ref{sec:Application}. This section contains a detailed explanation
for choosing nonthermal dark matter to illustrate the computations
of our paper instead of thermal dark matter. Finally, we summarize
our results in Section \ref{sec:Conclusions}. In appendices, we collect
technical details and also supplementary computational results: the
radiation transfer functions is derived in Appendix \ref{sec:Behaviors-of-Transfer},
a brief review of the gravitational Ward identity used for the gauge
invariance computation is given in Appendix \ref{sec:Review-of-Diffeomorphism},
the ADM formalism is reviewed in Appendix \ref{sec:ADM-formalism},
the details about the Pauli-Villars regulator is explained in Appendix
\ref{sec:Renormalization-of-Composite-Operators}, and the two point
function computation in the uniform curvature gauge is presented in
Appendix \ref{sec:Two-Point-Functions-in-UG}.

\section{A Class of Curvature and Isocurvature Perturbations\label{sec:Curvature-and-Isocurvature}}

Inflation through quantum correlator dynamics generates ``classical''
initial conditions for superhorizon cosmological fluid perturbations
\cite{Starobinsky:1980te,Guth:1982ec,Linde:1985yf,Kiefer:1998pb}.
The resulting initial conditions for the classical equations governing
classical fluid variables (which are set during radiation domination
before the CMB last scattering time) are categorized into two types:
adiabatic and isocurvature \cite{Linde:1984ti,Liddle:2000cg,Bucher:2004an,Bassett:2005xm}.
An adiabatic initial condition is intuitively characterized by all
species composing the fluid having the same initial number overdensities.
In the context of inflation, if there is a \emph{single dynamical
degree of freedom} $\phi$ during inflation such that after a few
efolds of inflation, the quantum vacuum boundary can be approximated
as Bunch-Davies initial conditions (for a discussion of number of
efold requirement see e.g. \cite{Chung:2003wn}), and if all the degrees
of freedom during radiation domination come from the inflaton decay,
then this adiabatic condition is the resulting approximate classical
boundary condition during radiation domination era of the universe.
An isocurvature initial condition intuitively corresponds to setting
nonzero the initial difference of the number overdensities of at least
one pair of fluid element species while setting to zero the total
energy density inhomogeneity on long wavelength scales. Because these
two types of initial conditions are linearly independent, a generic
initial condition to the linearized perturbation equations can be
written as a linear combination of them.

In this paper, we are concerned with the following physical system
which is generic for isocurvature scenarios. One real scalar slow-roll
inflaton degree of freedom $\phi$ dominates the energy density during
inflation. During this time period, there exists also another light
degree of freedom $\sigma$ which has no coupling to $\phi$ stronger
than gravity. We assume that this system carries an approximately
conserved discrete charge (such as $\mathbb{Z}_{2}$ broken at most
by a model dependent non-renormalizable operator) such that the one
particle states are stable and can act as dark matter. Note that since
we do not require all of the dark matter to come from $\sigma$, this
system is consistent with the existence of the weakly interacting
massive particle (WIMP) dark matter. If WIMP dark matter exists, the
parameter $\omega_{\sigma}\equiv\Omega_{\sigma}/\Omega_{CDM}<1$ will
play a role, and this scenario can yield interesting isocurvature
signatures for $\omega_{\sigma}$ as small as $10^{-5}$ \cite{Chung:2011xd}.
The action of this system can thus be written as 
\begin{equation}
S[\phi,\sigma,\{\psi\}]=\int(dx)\left\{ \frac{1}{2}M_{p}^{2}R+[-\frac{1}{2}g^{\mu\nu}\partial_{\mu}\phi\partial_{\nu}\phi-V(\phi)]+[-\frac{1}{2}g^{\mu\nu}\partial_{\mu}\sigma\partial_{\nu}\sigma-U(\sigma)]\right\} +S_{rh}[\phi,\{\psi\}]\label{eq:action_S-1}
\end{equation}
where $R$ is the Ricci scalar, $M_{p}^{2}=\frac{1}{8\pi G}$, $(dx)=d^{4}x\sqrt{|\det(g_{\mu\nu})|}$,
and $S_{rh}$ corresponds to the action of the reheating degrees of
freedom $\{\psi\}$. We assume that $\{\psi\}$ is heavy during inflation
such that it can be integrated out or if $\{\psi\}$ are light, they
are conformal such that they are not excited during inflation. After
inflation ends, we assume $\{\psi\}$ fields are light, leading to
a successful reheating scenario. The only special initial condition
dependent assumption that we make in this isocurvature scenario is
that $\langle\sigma\rangle\ll H/(2\pi)$ during inflation even when
$\partial^{2}U(\sigma)/\partial\sigma^{2}\ll H$. %
\footnote{Note that even with a Gaussian distributed values of $\langle\sigma\rangle$
on an inflationary patch with a Gaussian width $H/(2\pi)$, there
is about a 2/3 probability that such initial condition configurations
can be found. Also, an unbroken discrete symmetry such as $\mathbb{Z}_{2}:\sigma\rightarrow-\sigma$
can stabilize the VEV. In the context of supergravity, generic terms
in the effective potential however can appear leading to $\langle\sigma\rangle\neq0$
during inflation. In the end, whether or not $\langle\sigma\rangle=0$
is model dependent, but it is not fine tuned.%
} ~~Because $\langle\sigma\rangle=0$ during inflation, $\sigma$
by itself does not spontaneously break time translation invariance
and therefore does not mix with $\delta\phi$ in forming the gauged
time translation Nambu-Goldstone boson $\zeta$. Hence, we can treat
the scalar fluid variable $\zeta(\delta g_{\mu\nu},\delta\phi)$ as
the curvature degree of freedom and $\delta_{S}(\sigma,\zeta)$ as
the isocurvature degree of freedom. (As we will show in detail below,
the isocurvature degree of freedom $\delta_{S}$ will be quadratic
in $\sigma$ and will involve $\zeta$ as a difference). 

Thus, the basic physics picture of the classical fluid that we are
concerned with in this paper is the following. To predict CMB temperature
fluctuation $\langle\Delta T\Delta T\rangle$, we must compute the
cross correlation $\langle\delta_{S}\zeta\rangle$ since at the linearized
level, Einstein-Boltzmann equations give the relationship $\Delta T/T\sim c_{1}\zeta+c_{2}\delta_{S}$
for computable order unity (for long wavelengths) coefficients $c_{i}$.
Up until this paper, there has never been an explicit computation
of the $\langle\delta_{S}\zeta\rangle/\sqrt{\langle\zeta\zeta\rangle\langle\delta_{S}\delta_{S}\rangle}$
coming from irreducible gravitational interactions.%
\footnote{As we will later explain, we do not compute $\langle\delta_{S}\zeta\rangle$
analytically fully beyond the time of the end of inflation. However,
the importance of the isocurvature cross correlation can be generically
predicted by $\langle\delta_{S}\zeta\rangle/\sqrt{\langle\zeta\zeta\rangle\langle\delta_{S}\delta_{S}\rangle}$
which is insensitive to the post-inflationary evolution for superhorizon
modes.%
} What will emerge is a clean universal result that applies to a wide
range of isocurvature models including those of the QCD axions (in
a particular initial condition regime) and WIMPZILLAs. We find that
$\langle\delta_{S}\zeta\rangle$ contribution is generically subdominant
to $\langle\delta_{S}\delta_{S}\rangle$ in the case of pure gravitational
interactions.

In the following, we establish our conventions in describing this
isocurvature degree of freedom carrying the non-adiabatic initial
condition information. In the process, we review the gauge invariant
construction of these cosmological perturbations and the current CMB
observational constraint, which represents the strongest constraint
on the isocurvature initial condition derived from inflation.

\subsection{Gauge Invariant Construction\label{sub:Gauge-Invariant-Construction}}

The cosmological inhomogeneity perturbation variables are generally
spacetime coordinate gauge-dependent because of the coordinate dependent
definition of fictitious background metric slices. From the perspective
of matching classical equation initial conditions to inflationary
quantum correlator computations, identifying gauge invariant combinations
is helpful \cite{Bardeen:1980kt,Mukhanov:1990me,Weinberg:2008zzc}.
On the other hand, the gauge freedom involved in computing gauge invariant
quantities facilitates the quantum computation. Hence, understanding
the gauge dependences of the correlation computations is helpful.
In this subsection, we review the gauge invariant variable construction
and establish our notation. For a more general discussion, see for
example \cite{Bardeen:1980kt,Kodama:1985bj,Mukhanov:1990me,Bruni:1996im,Wands:2000dp,Prokopec:2012ug,Arroja:2011yj,Langlois:2010vx,Rigopoulos:2011eq,Malik:2008im,Acquaviva:2002ud,Ellis:1989jt,Hwang:1991aj}.

In $(t,\vec{x})$ coordinates, we parameterize the metric as $g_{\mu\nu}=\bar{g}_{\mu\nu}+\delta g_{\mu\nu}^{(S)}$
where the scalar metric perturbation is

\begin{equation}
\delta g_{\mu\nu}^{(S)}=\left(\begin{array}{cc}
-E & aF_{,i}\\
aF_{,i} & a^{2}[A\delta_{ij}+B_{,ij}]
\end{array}\right),\label{eq:deltag_s}
\end{equation}
the background metric is $\bar{g}_{\mu\nu}\equiv\mbox{\mbox{diag}}\{-1,a^{2}(t),a^{2}(t),a^{2}(t)\},$
and derivatives are denoted as usual as $X_{,i}\equiv\partial X/\partial x^{i}$.
Under the diffeomorphism $x\rightarrow x+\epsilon$ where 
\begin{equation}
\epsilon^{\mu}=(\epsilon^{0},a^{-2}\partial_{i}(\epsilon^{S})),\label{eq:diffeodef}
\end{equation}
the scalar metric perturbation components transform as 
\begin{eqnarray}
\Delta A & = & -2H\epsilon^{0},\,\,\,\,\,\,\,\,\Delta B=-\frac{2}{a^{2}}\epsilon^{S},\\
\Delta E & = & -2\dot{\epsilon}^{0},\,\,\,\,\,\,\,\,\Delta F=\frac{1}{a}(\epsilon^{0}-\dot{\epsilon}^{S}+2H\epsilon^{S})
\end{eqnarray}
which is obtained from $\delta g_{\mu\nu}^{(S)}\rightarrow\delta g_{\mu\nu}^{(S)}+\Delta(\delta g_{\mu\nu}^{(S)})$
with $\Delta(\delta g_{\mu\nu}^{(S)})=-\mathcal{L}_{\epsilon^{\mu}\partial_{\mu}}\bar{g}_{\mu\nu}$. 

Similarly, we parameterize the perfect fluid stress tensor for a fluid
element $a$ as 
\begin{equation}
T_{\mu\nu}^{(a)}=\bar{T}_{\mu\nu}^{(a)}+\delta T_{\mu\nu}^{(a)}
\end{equation}
where $\bar{T}_{\mu\nu}^{(a)}\equiv\mbox{diag}\{\bar{\rho}_{a},\bar{P}_{a},\bar{P}_{a},\bar{P}_{a}\}$
contains the average energy density and pressure seen by a comoving
observer, $\delta T_{ij}^{(a)}=\bar{P}_{a}\delta g_{ij}^{(S)}+a^{2}\delta_{ij}\delta P_{a}$,
$\delta T_{i0}^{(a)}=\bar{P}_{(a)}\delta g_{i0}^{(S)}-(\bar{\rho}_{a}+\bar{P}_{a})\delta U_{i}^{(a)}$
(where $\delta U_{i}^{(a)}$ is the velocity perturbation), and $\delta T_{00}^{(a)}=-\bar{\rho}_{a}\delta g_{00}^{(S)}+\delta\rho_{a}$.
Under the diffeomorphism of Eq.~(\ref{eq:diffeodef}), the energy
density perturbation transforms as 
\begin{equation}
\Delta\delta\rho_{a}=-\epsilon^{0}\dot{\bar{\rho}}_{a}.
\end{equation}

In practice, gauge-invariant variables are constructed by combining
metric perturbations and other perturbations, such as densities. A
popular choice is 
\begin{equation}
\zeta_{a}\equiv\frac{A}{2}-H\frac{\delta\rho_{a}}{\dot{\bar{\rho}}_{a}}.
\end{equation}
For example, the first-order gauge-invariant perturbation associated
with the inflaton $\phi$ is usually defined as 

\begin{eqnarray}
\zeta_{\phi} & \equiv & \frac{A}{2}-H\frac{\delta\rho_{\phi}}{\dot{\bar{\rho}}_{\phi}}\label{eq:inflatonzeta}
\end{eqnarray}
(see for example Ref. \cite{Weinberg:2008zzc} and references therein).
Now, one can form a quantity that is conserved through reheating by
defining
\begin{equation}
\zeta_{\mbox{tot}}\equiv\sum_{i}r_{i}\zeta_{i}\label{eq:zetatotdef}
\end{equation}
where
\begin{equation}
r_{i}\equiv\frac{\bar{\rho}_{i}+\bar{P}_{i}}{\sum_{n}\bar{\rho}_{n}+\bar{P}_{n}}.\label{eq:ratiodef}
\end{equation}
Because there must be reheating dynamical degrees of freedom, $\zeta_{\mbox{tot}}$
must involve at least 2 degrees of freedom by the end of inflation
of any single field slow-roll model. In single field slow-roll scenarios,
what is done in practice is to argue that the reheating degrees of
freedom are integrated out during inflation and then integrated back
in at the end of inflation due to the different location of the inflaton
VEV at the end of inflation. Alternatively, another often used assumption
is that the main reheating degree of freedom are conformal such that
no isocurvature fluctuations are appreciably excited during inflation.
This means that in single field models, we have
\begin{equation}
\zeta_{\mbox{tot}}\approx\zeta_{\phi}\label{eq:zetatotiszeta}
\end{equation}
up to ambiguities in how one hides the reheating degrees of freedom. 

One reason why the combination of Eq\@.~(\ref{eq:zetatotdef}) is
convenient is because the superhorizon mode of this is approximately
conserved through reheating if this mode object can be shown to obtain
an initial conditions of what is sometimes referred to as the adiabatic
solution \cite{Weinberg:2003sw,Weinberg:2008zzc} and there are no
non-adiabatic processes that mix superhorizon modes of isocurvature
degrees of freedom with $\zeta_{\mbox{tot}}$. Such classical adiabatic
solution initial conditions are generated by the Bunch-Davies quantum
fluctuations for $\zeta_{\phi}$, and we will restrict the couplings
of the isocurvature degrees of freedom (discussed below) such as to
avoid non-adiabatic mixing. This means that Eq.~(\ref{eq:zetatotiszeta})
ensures that $\zeta_{\mbox{tot}}$ is approximately conserved if $\bar{\rho}_{\phi}+\bar{P}_{\phi}$
dominates over others. More explicitly, as discussed in the introduction
to this section, suppose there exists only one isocurvature field
degree of freedom which we call $\sigma$ during the inflationary
period.%
\footnote{The species $\sigma$ will later be identified dark matter candidates
such as the axions and WIMPZILLAs.%
} The total curvature perturbation can be written as 
\begin{equation}
\zeta_{\mbox{tot}}=\zeta_{\phi}+r_{\sigma}(\zeta_{\sigma}-\zeta_{\phi})\label{eq:includingrsig}
\end{equation}
with the sum over $n$ runs over $\phi$ and $\sigma$ (assuming that
$\psi$ has been integrated out during inflation). However, one can
estimate that the coefficient of $\zeta_{\sigma}$ during inflation
is 
\begin{equation}
r_{\sigma}\lesssim\frac{1}{(2\pi)^{2}}\Delta_{\zeta}^{2}\sim10^{-11}\label{eq:smallrsig}
\end{equation}
which makes the approximation of $\zeta_{\mbox{tot}}\approx\zeta_{\phi}$
accurate, just as in the single field case of Eq.~(\ref{eq:zetatotiszeta}).
Thus just as in the single field scenarios without $\sigma$, $\zeta_{\mbox{tot}}$
acquires an approximately adiabatic boundary condition from the Bunch-Davies
vacuum field fluctuations.

To complete the examination of how $\zeta_{\mbox{tot}}$ is used in
the scenario of concern in this paper, let's look at the time period
surrounding the reheating transition when the universe reaches radiation
domination. Near the time of the completion of the reheating, the
variable $\zeta_{\mbox{tot }}$ is approximately 
\begin{equation}
\zeta_{\mbox{tot}}\approx r_{\phi}\zeta_{\phi}+\sum_{i}r_{\psi_{i}}\zeta_{\psi_{i}}\label{eq:approxzetatot}
\end{equation}
such that after the inflaton decays, we have $r_{\phi}=0$ and 
\begin{equation}
\zeta_{\mbox{tot}}\approx\sum_{i}r_{\psi_{i}}\zeta_{\psi_{i}}.
\end{equation}
\footnote{In the case that $\psi_{i}$ is integrated back in at the end of inflation,
we have made the assumption that this does not change $\zeta_{\mbox{tot}}$%
} (The approximation used in Eq.~(\ref{eq:approxzetatot}) neglects
the $r_{\sigma}$ contribution because of Eq.~(\ref{eq:smallrsig}).)
It is also a standard assumption that 
\begin{equation}
\zeta_{\psi_{i}}=\zeta_{\mbox{tot}},\label{eq:radzetatot}
\end{equation}
which is rigorously true if one relativistic species dominate the
fluid (e.g. $r_{\psi_{1}}\approx1$) or if the decay process does
not redistribute the spatial inhomogeneities of $\psi_{i}$ in a distinct
configuration from that of $\phi$. %
\footnote{However this need not be true for more general reheating scenarios.%
} This justifies the usual statement in the literature that $\zeta_{\mbox{tot}}$
defined in Eq.~(\ref{eq:zetatotdef}) is primarily useful for arguing
how a combination of quantities involving the inflaton and the reheating
decay products remain unchanged through the reheating phase transition.
Here, we have merely described how this argument is not changed by
the presence of $\sigma$ because of the smallness of $r_{\sigma}$
in Eq.~(\ref{eq:smallrsig}) during the primordial periods of interest.

In summary, as long as boundary conditions for the classical fluid
equation are evaluated at a time when $r_{\sigma}$ is small (compared
to the accuracy desired), we can neglect the $r_{\sigma}$ contribution
from $\zeta_{\mbox{tot}}$ both through reheating and until the time
that boundary conditions for the classical fluid equations are imposed.
Hence, if $\zeta_{\mbox{tot}}$ remains constant on long wavelengths
(due to the initial conditions set by the Bunch-Davies vacuum), Eqs.~(\ref{eq:includingrsig})
and (\ref{eq:smallrsig}) imply that the effective curvature perturbation
during this early primordial epoch is given by Eq.~(\ref{eq:zetatotiszeta}).
Hence, in the discussion below, we will drop the $\phi$ subscript
and write
\begin{equation}
\zeta\equiv\zeta_{\phi}\approx\zeta_{\mbox{tot}}.\label{eq:zetadefined}
\end{equation}
During this radiation dominated early primordial time $t_{p}$, the
relationship between super horizon $A(t_{p},\vec{k})$ and the value
of $\zeta(t_{e},\vec{k})$ evaluated at the end of inflation time
$t_{e}$ is
\begin{equation}
\frac{A(t_{p},\vec{k})}{2}\approx\frac{2}{3}\zeta(t_{e},\vec{k})
\end{equation}
in the Newtonian gauge ($B=F=0$) and the presence of $\zeta_{\sigma}$
gives a small error controlled by $r_{\sigma}$.

At the same radiation dominated era%
\footnote{During this time period, there is possibly a population of thermal
dark matter components such as thermal WIMPs.%
} when initial condition is set by $\zeta_{\mbox{tot}}\approx\zeta$,
the inhomogeneity of the small mixture of dark matter component $\sigma$
can be related to the isocurvature perturbation $\zeta_{\sigma}$.
Conventionally, this information is parameterized by the gauge-invariant
isocurvature perturbation \cite{Wands:2000dp,Liddle:2000cg,Bassett:2005xm}
\begin{equation}
\delta_{S}(t,\vec{k})\equiv3\left(\zeta_{\sigma}(t,\vec{k})-\zeta_{\mbox{tot}}(t,\vec{k})\right).\label{eq:deltaS}
\end{equation}
The physical interpretation of this quantity can be see by noting
that when $\sigma$ particles are dominantly non-relativistic and
the universe is radiation dominated, this expression becomes
\begin{equation}
\delta_{S}(t,\vec{k})=\frac{\delta\rho_{\sigma}(t,\vec{k})}{\bar{\rho}_{\sigma}}-\frac{3}{4}\frac{\delta\rho_{\gamma}(t,\vec{k})}{\bar{\rho}_{\gamma}}
\end{equation}
where $\rho_{\gamma}$ represents the photon energy densities. This
clearly represents the difference in number densities of $\sigma$
and $\gamma$.%
\footnote{It is interesting to note that since number densities can diverge
while gravitational physics does not care about number densities (in
favor of energy densities), this choice of variables is unfortunate
in situations when there are IR divergences. In this paper, we stick
to this convention which is prevalent in literature.%
} Assuming that the radiation inhomogeneity is characterized by $\zeta$
as explained in Eqs.~(\ref{eq:radzetatot}) and (\ref{eq:zetadefined})
during radiation domination, we have
\begin{equation}
\delta_{S}(t,\vec{k})\approx3(\zeta_{\sigma}(t,\vec{k})-\zeta(t_{e},\vec{k}))
\end{equation}
Similarly to the case of $\zeta_{\mbox{tot}}$, long wavelength limit
of $\zeta_{\sigma}$ generated from Bunch-Davies initial conditions
simplify (partly because of causality) in the absence of non-adiabatic
processes mixing of $\zeta_{\mbox{\ensuremath{\sigma}}}$ with other
superhorizon degrees of freedom. The $\zeta_{\sigma}$ mode for a
comoving wave vector $\vec{k}$ becomes constant once $|\vec{k}/a|\ll H$
and $m_{\sigma}\ll H$ because the mode functions involved in $\zeta_{\sigma}$
are governed by the Hubble friction once these conditions are satisfied.

Although the key correlator computation result of this paper involving
$\beta$ evaluated at the end of inflation is independent of the transfer
function evolving the isocurvature degrees of freedom after the end
of inflation, because its immediate phenomenological application to
CMB requires a transfer function describing this post-inflationary
evolution, we will restrict our illustration in Section \ref{sec:Application}
to the situation when the chemical reaction rates that mix $\sigma$
and the radiation components are negligible. We will discuss in more
detail the cross section constraint for this condition in Appendix
\ref{sub:Weakness}.

\subsection{Observational Constraints on Isocurvature Perturbation\label{sub:Observational-Constraints-on-Isocurvature}}

The current observational data shows that the CMB power spectrum is
consistent with the adiabatic initial conditions. However, it does
not rule out mixed boundary condition contributions from CDM isocurvature
perturbations. Schematically, the temperature fluctuations depend
linearly on $\zeta$ and $\delta_{S}$ initial conditions as 
\begin{equation}
\frac{\Delta T}{T}=c_{1}\zeta+c_{2}\delta_{S}
\end{equation}
where $c_{i}\sim O(1)$. Hence, the CMB temperature correlation data
constrains
\begin{equation}
\frac{k^{3}}{2\pi^{2}}\int\frac{d^{3}p}{(2\pi)^{3}}\langle\frac{\Delta T(\vec{p})}{T}\frac{\Delta T^{*}(\vec{k})}{T}\rangle=\Delta_{\zeta}^{2}(k)\left[|c_{1}|^{2}+|c_{2}|^{2}\frac{\alpha}{1-\alpha}-2\Re\left(c_{1}^{*}c_{2}\beta\sqrt{\frac{\alpha}{1-\alpha}}\right)\right]\label{eq:physicalroleofbeta}
\end{equation}
where \cite{Bean:2006qz}
\begin{equation}
\int\frac{d^{3}p}{(2\pi)^{3}}\langle\zeta(\vec{p})\zeta^{*}(\vec{k})\rangle=\Delta_{\zeta}^{2}(k)\frac{2\pi^{2}}{k^{3}}
\end{equation}
\begin{equation}
\int\frac{d^{3}p}{(2\pi)^{3}}\langle\delta_{S}(\vec{p})\delta_{S}^{*}(\vec{k})\rangle=\Delta_{\delta_{S}}^{2}(k)\frac{2\pi^{2}}{k^{3}}
\end{equation}
\begin{equation}
\int\frac{d^{3}p}{(2\pi)^{3}}\langle\delta_{S}(\vec{p})\zeta^{*}(\vec{k})\rangle=\Delta_{\zeta\delta_{S}}^{2}(k)\frac{2\pi^{2}}{k^{3}}
\end{equation}
\begin{eqnarray}
\alpha & \equiv & \frac{\Delta_{\delta_{S}}^{2}(k)}{\Delta_{\zeta}^{2}(k)+\Delta_{\delta_{S}}^{2}(k)},\label{eq:fractional_powerspectrum}\\
\beta & \equiv & -\frac{\Delta_{\zeta\delta_{S}}^{2}(k)}{\sqrt{\Delta_{\zeta}^{2}(k)\Delta_{\delta_{S}}^{2}(k)}},\label{eq:fractional_cross-correlation}
\end{eqnarray}
\footnote{Our sign conventions are such that \emph{negative} values for $\beta$
correspond to a positive contribution of the cross-correlation term
to the Sachs-Wolfe component of the total temperature spectrum. See,
e.g., \cite{Komatsu:2008ex,Komatsu:2010fb}.%
}which are customarily evaluated in the primordial epoch when $k$
corresponds to a far superhorizon scale such that the $\Delta_{X}^{2}(k)$
objects are constant in time. Typically the data constraints are parameterized
by evaluating $\alpha$ and $\beta$ at a pivot scale $k=k_{0}$ \cite{Komatsu:2008ex,Komatsu:2010fb}.
An important utility of this parameterization is the following fact:
a necessary and sufficient condition for the cross correlation to
be a significant part of the isocurvature contribution is to have
$\left|\beta\right|\gtrsim|c_{2}/c_{1}|\sqrt{\alpha}$ for $\alpha<1$.
For example, in order to have approximately the same level of the
angular power spectra from both pure isocurvature correlation and
and cross-correlation at the intermediate scale $l\sim200$, i.e.
$C_{l}^{pure\, iso}\sim C_{l}^{cross\, cor}$, the fractional cross-correlation
should satisfy $\left|\beta\right|\gtrsim4\times10^{-2}$. Another
utility of the $\beta$ variable comes from the fact that when there
are non-trivial transfer functions governing $\Delta_{\zeta\delta_{S}}^{2}$
and $\Delta_{\delta_{S}}^{2}$ after the end of inflation, the transfer
function factors can cancel in the expression for $\beta$. We will
use this feature later to compute $\beta$ based on just the (quasi)-dS
mode function behavior.%
\footnote{We will use the exact dS approximation for the massive $\sigma$ and
use the quasi-dS approximation for only the massless scenario. The
corrections coming from the the deviations away from the exact dS
background in principle can be absorbed into the transfer function
multiplying the superhorizon mode function which cancel out in $\beta$
due to a common appearance in the numerator and the denominator.%
} 

\begin{figure}
\begin{centering}
\includegraphics{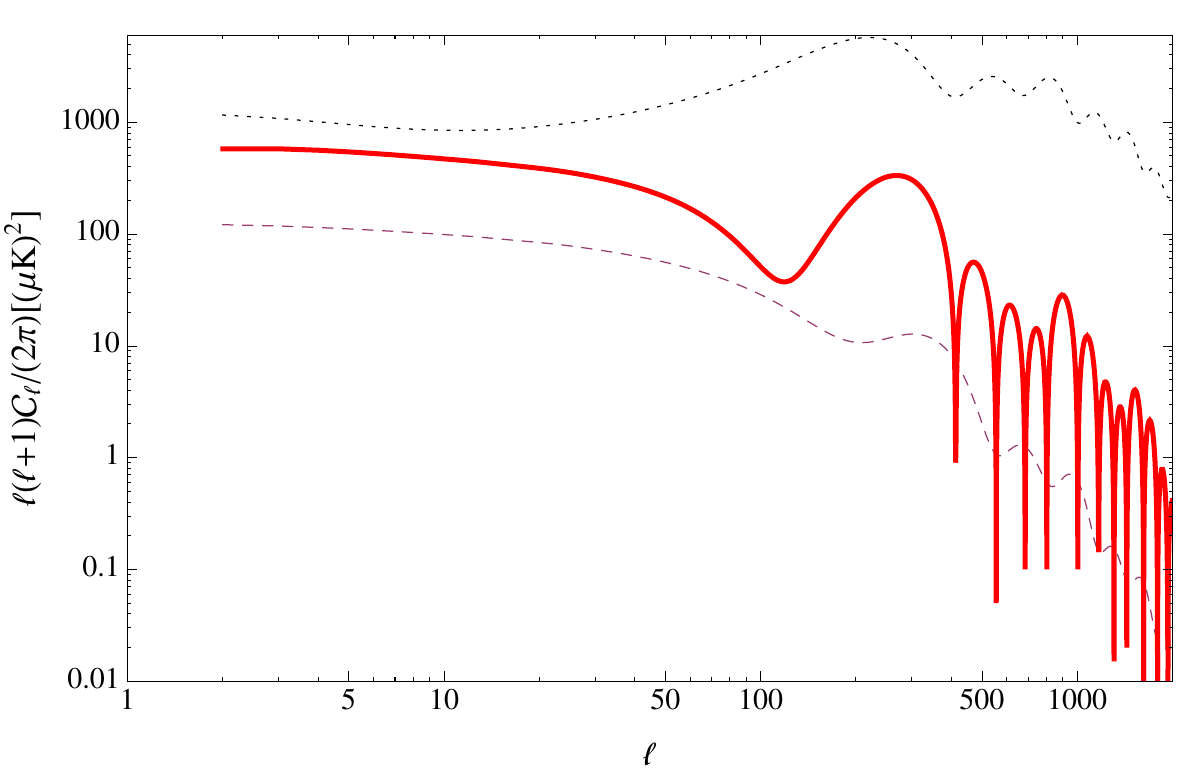}
\par\end{centering}

\caption{\label{fig:Powerspectrum}Angular power spectra $C_{l}$ from pure
adiabatic(dotted), pure isocurvature(dashed) contributions. The solid
red line corresponds to be qualitatively the contribution of the absolute
magnitude of the third term in Eq. (\ref{eq:physicalroleofbeta}).
The plotted pure adiabatic perturbation has the spectral index $n_{s}=0.96$.
For isocurvature perturbations, the spectral index $n_{iso}$ is 1
and the isocurvature fraction $\alpha=0.067$ defined at $k_{0}=0.002\, Mpc^{-1}$,
and the fractional cross-correlation $\left|\beta\right|$ is 1.}
\end{figure}

As far as the experimental numbers are concerned, the isocurvature
contribution to the CMB temperature perturbation is expected to be
roughly less than 10\% compared to the curvature contribution. More
precisely, the Planck+WP limits \cite{Komatsu:2010in,Hinshaw:2012fq,Ade:2013rta}
are 
\begin{equation}
\alpha|_{\beta=0}<0.016\mbox{ (95\% CL) and }\alpha|_{\beta=-1}<0.0011\mbox{ (95\% CL)},
\end{equation}
where the isocurvature power spectrum is assumed to be scale-invariant,
i.e. $n_{iso}=1$. The significant difference in the upper-bound of
$\alpha$ between uncorrelated and totally (anti-)correlated cases
can be explained by the ratio $\beta/\sqrt{\alpha}$ already discussed
above. The difficulty in improving the current isocurvature bound
with data on short wavelengths can be seen in Fig. \ref{fig:Powerspectrum},
where one sees a fall-off of the isocurvature spectrum on short scales
$(l\gtrsim100$). This fall-off is generic and can be attributed to
the transfer function effect encoded by $c_{1}(k)/c_{2}(k)$ in Eq.~(\ref{eq:physicalroleofbeta})
for $k\gtrsim k_{eq}$ (where $k_{eq}/a_{0}\sim10^{-2}\mbox{ Mpc}^{-1}$
is the wave vector associated with matter radiation equality). To
understand why $c_{1}(k)/c_{2}(k)$ generically becomes large for
$k\gtrsim k_{eq}$, note that isocurvature modes with $k\gtrsim k_{eq}$
enter the horizon during radiation domination. Because the isocurvature
effect on the temperature spectrum is gravitational, the value of
$c_{1}(k)/c_{2}(k)$ is proportional to the ratio $\rho_{R}(t(k))/\rho_{\sigma}(t(k))$
of the radiation energy density to the energy density in the isocurvature
degree of freedom at the time $t(k)$ when mode $k\gtrsim k_{eq}$
enters the horizon. Since shorter wavelengths enter the horizon earlier,
$\rho_{R}(t(k))/\rho_{\sigma}(t(k))$ is larger for shorter wavelengths,
making $c_{1}(k)/c_{2}(k)$ larger. For those readers not familiar
with this physics, some of the details of the transfer function are
reviewed in Appendix \ref{sec:Behaviors-of-Transfer}.

Because of the large differences in the constraints between $\beta=0$
and $\beta=-1$, estimating the cross-correlation is crucial to restrict
parameters and give observable predictions of isocurvature models.
In particular, the axion scenario with a negligible homogeneous vacuum
misalignment angle (and similarly the WIMPZILLA scenario with a negligible
homogenous background field value) predicts detectable non-Gaussianity
\cite{Hikage:2008sk,Kawasaki:2008sn,Chung:2011xd} 
\begin{equation}
f_{NL}\sim30\left(\frac{\alpha}{0.067}\right)^{3/2}\label{eq:fnl}
\end{equation}
provided the assumption the cross-correlation is zero, i.e. $\beta=0$.
However, as we will explain, this assumption is not obvious for massive
field quadratic isocurvature scenarios, and the reexamination of this
assumption is one of the goals of this paper.

\section{Computation of Correlators\label{sec:Computation-of-Correlators}}

In order to provide the initial condition of the classical fluid equations,
it is standard to compute the quantum equal time correlators with
the inflationary background approximated as a Bunch-Davies vacuum.
In this section, we compute the correlators using the ``in-in''
formalism (e.g. see Weinberg \cite{Weinberg:2005vy}). More specifically,
in the context of canonical quantization, we perturbatively compute
the expectation value of an operator $\hat{Q}(t)$
\begin{equation}
\left\langle \hat{Q}(t)\right\rangle =\sum_{n}\left(-i\right)^{n}\int_{-\infty}^{t}dt_{1}\int_{-\infty}^{t_{1}}dt_{2}\cdots\int_{-\infty}^{t_{n-1}}dt_{n}\left\langle \left[\left[\left[\hat{Q}^{I}(t),\hat{H}^{I}(t_{n})\right],\hat{H}^{I}(t_{n-1})\right],\cdots\hat{H}^{I}(t_{1})\right]\right\rangle ,\label{eq:in-in_formalism}
\end{equation}
where the superscript $I$ stands for the interaction picture and
$\hat{Q}(t)$ represents a product of canonically quantized operators.

In the scenario explained in Sec.~\ref{sec:Curvature-and-Isocurvature},
we consider the gravitational coupling whose interaction Hamiltonian
is derived from the ADM formalism with a given choice of gauge. For
the computation of the cross-correlation to leading order in gravitational
coupling, we need at least up to the cubic coupling $H_{\zeta\sigma\sigma}^{I}$,
where $\sigma$ is a spectator field during inflation. The interaction
Hamiltonian is diffeomorphism gauge-dependent. For two commonly used
gauges, the comoving gauge$\ensuremath{(\delta\phi=0)}$ and the uniform
curvature gauge$(A=0)$, we have
\begin{eqnarray}
H_{\zeta\sigma\sigma}^{I}(t) & = & -\frac{1}{2}\int d^{3}x\, a^{3}(t)\, T_{\sigma}^{\mu\nu}(t,\vec{x})\delta g_{\mu\nu}(t,\vec{x}),\label{eq:Hint}\\
\delta g_{\mu\nu}^{(C)} & = & \left(\begin{array}{cc}
-2\frac{\dot{\zeta}}{H} & (-\frac{\zeta}{H}+\epsilon\frac{a^{2}}{\nabla^{2}}\dot{\zeta})_{,i}\\
(-\frac{\zeta}{H}+\epsilon\frac{a^{2}}{\nabla^{2}}\dot{\zeta})_{,i} & a^{2}\delta_{ij}2\zeta
\end{array}\right),\label{eq:deltag_CG}\\
\delta g_{\mu\nu}^{(U)} & = & \left(\begin{array}{cc}
2\epsilon\zeta & \epsilon\frac{a^{2}}{\nabla^{2}}\dot{\zeta}_{,i}\\
\epsilon\frac{a^{2}}{\nabla^{2}}\dot{\zeta}_{,i} & 0
\end{array}\right),
\end{eqnarray}
where $T_{\sigma}^{\mu\nu}$ is the stress energy tensor of the field
$\sigma$, and $\delta g_{\mu\nu}$ is the metric perturbation and
the superscript $(C)$ and $(U)$ denote the comoving gauge and uniform
curvature gauge, respectively. A detailed derivation of the interaction
Hamiltonian using the ADM formalism is presented in Section \ref{sec:ADM-formalism}.

The isocurvature perturbation $\delta_{S}$ should be also written
in terms of quantum operators associated with the energy density $\rho_{\sigma}$
of the particle $\sigma$. Since the energy density $\rho_{\sigma}$
is written in bilinear form of $\sigma$ and since the energy density
of CDM are often those of non-relativistic particles at the time of
matching to classical equations, we may approximate the energy density
$\rho_{\sigma}\approx m_{\sigma}^{2}\sigma^{2}$. We then promote
field $\sigma$ to a quantum operator:
\begin{equation}
\delta_{\sigma}\equiv\frac{\delta\rho_{\sigma}}{\rho_{\sigma}}\approx\frac{\sigma^{2}-\bar{\sigma}^{2}}{\bar{\sigma}^{2}}\to\hat{\delta}_{\sigma}=\frac{\hat{\sigma}^{2}-\left\langle \hat{\sigma}^{2}\right\rangle }{\left\langle \hat{\sigma}^{2}\right\rangle }.
\end{equation}
The field $\hat{\sigma}$ can be decomposed into the classical homogeneous
background and the quantized perturbation, i.e. $\hat{\sigma}=\bar{\sigma}+\delta\hat{\sigma}$.
Unlike the inflaton $\phi$ whose classical background is non-zero,
because we consider the field $\hat{\sigma}$ without classical background,
the leading density perturbation starts with the quadratic in the
operator $\delta\hat{\sigma}^{2}$. As with any quantum composite
operator, we renormalize it with counter terms invariant under the
underlying gauge symmetry (here, it is diffeomorphism): 
\begin{equation}
\left(\hat{\sigma}^{2}\right)_{r}=\left(\delta\hat{\sigma}+\sum_{i}\hat{\chi}_{i}\right)^{2}+\delta Z_{0}+\delta Z_{1}R,
\end{equation}
where the subscript $r$ denotes that the operator is a renormalized
composite operator, $R$ is the Ricci scalar, and $\hat{\chi}_{i}$
are Pauli-Villars fields, which is described in Section \ref{sec:Renormalization-of-Composite-Operators}.
We apply this to gauge-invariant isocurvature variable $\delta_{S}$
defined in Section \ref{sub:Gauge-Invariant-Construction}. Then we
have
\begin{eqnarray}
\hat{\delta_{S}}^{(C)} & = & -\frac{3H}{\partial_{t}\langle\left(\hat{\sigma}^{2}\right)_{r}\rangle}\left[\left(\hat{\sigma}^{2}\right)_{r}-\langle\left(\hat{\sigma}^{2}\right)_{r}\rangle\right],\\
\hat{\delta_{S}}^{(U)} & = & -\frac{3H}{\partial_{t}\langle\left(\hat{\sigma}^{2}\right)_{r}\rangle}\left[\left(\hat{\sigma}^{2}\right)_{r}-\langle\left(\hat{\sigma}^{2}\right)_{r}\rangle\right]-3\hat{\zeta}.\label{eq:deltaS_UG}
\end{eqnarray}
We will not write the hat explicitly from now on.

In the next subsection, we present how a non-diffeomorphism-invariant
estimation of the cross-correlation leads to an observationally attractive
but grossly incorrect result. In subsections after that, we identify
the problems with the wrong estimate and calculate the cross-correlation
properly.

\subsection{Plausible but Wrong Estimation of the Cross-Correlation\label{sub:Puzzles}}

In this subsection, we present a plausible estimation of the cross
correlation that leads to a large value that is observationally interesting.
Unfortunately, we will see in later subsections that the estimate
presented in this subsection can be many orders of magnitude off due
to the explicit breaking of diffeomorphism invariance in the treatment
of the UV physics. Nonetheless, what is presented in this subsection
is interesting both as a lesson in field theory and as a motivation
for the careful correct computation that follows later. 

The isocurvature cross-correlation in the comoving gauge is written
as
\begin{equation}
\left\langle \delta_{S}^{(C)}\zeta\right\rangle \approx\frac{\left\langle \left(\sigma^{2}\right)_{r}\zeta\right\rangle }{\left\langle \left(\sigma^{2}\right)_{r}\right\rangle },
\end{equation}
where we have used $\partial_{t}\left\langle \left(\sigma^{2}\right)_{r}\right\rangle +3H\left\langle \left(\sigma^{2}\right)_{r}\right\rangle \approx0$
for the isocurvature field number density. For an order of magnitude
estimation, we consider a non-derivatively coupled part of the gravitational
interaction, $2\zeta a^{2}\delta_{ij}T_{\sigma}^{ij}\in H_{\zeta\sigma\sigma}^{I}$.
Then the two-point function, shown diagrammatically in Fig. \ref{fig:Two-point-function},
is written in the Fourier space as 
\begin{eqnarray}
\widetilde{\left\langle \left(\sigma^{2}\right)_{r}\zeta\right\rangle _{p}^{C}} & \sim & \int d^{3}x\, e^{-i\vec{p}\cdot\vec{x}}\int^{t}d^{4}z\, a^{3}(t_{z})\left\langle \left[\sigma^{2}(t,\vec{x})\zeta(t,\vec{0}),\frac{i}{2}\left(2\zeta a^{2}\delta_{ij}T_{\sigma}^{ij}\right)_{z}\right]\right\rangle \label{eq:dummyeq}\\
 & \sim & -4\int\frac{d^{3}k_{1}}{\left(2\pi\right)^{3}}d^{3}k_{2}\delta^{3}(\vec{k}_{1}+\vec{k}_{2}-\vec{p})\int_{-\infty}^{t}dt_{z}\, a_{z}^{3}\nonumber \\
 &  & \times\mbox{Im}\left[\zeta_{p}(t)\zeta_{p}^{*}(t_{z})u_{k_{1}}(t)u_{k_{2}}(t)\left\{ \frac{1}{2}\frac{\vec{k}_{1}\cdot\vec{k}_{2}}{a^{2}}+3\left(\frac{1}{2}\partial_{t}^{(1)}\partial_{t}^{(2)}-\frac{1}{2}m_{\sigma}^{2}\right)\right\} u_{k_{1}}^{*}(t_{z})u_{k_{2}}^{*}(t_{z})\right]
\end{eqnarray}
where
\begin{equation}
\widetilde{\left\langle AB\right\rangle _{p}}\equiv\int d^{3}x\, e^{-i\vec{p}\cdot\vec{x}}\left\langle A(t,\vec{x})B(t,0)\right\rangle ,
\end{equation}
$\zeta_{p}$ and $u_{k}$ are mode functions for $\zeta$ and $\sigma$,
respectively, and $\partial_{t}^{(i)}$ means the time derivative
with respect to $u_{k_{i}}^{*}(t_{z})$.%
\footnote{It is also helpful to remember that in terms of Fourier space operators/fields,
the tilde notation is equivalent to
\[
\widetilde{\left\langle AB\right\rangle _{p}}=\int\frac{d^{3}p_{2}}{(2\pi)^{3}}\left\langle A(t,\vec{p})B(t,\vec{p}_{2})\right\rangle 
\]
where
\[
A(t,\vec{p})\equiv\int d^{3}xe^{-i\vec{p}\cdot\vec{x}}A(t,\vec{x})
\]
for generic operators/fields $A$ and $B$.%
} This integral is UV divergent, and thus we introduce the horizon
scale UV cut-off 
\begin{equation}
\Lambda_{UV}\sim aH_{inf}.\label{eq:uvcutoff}
\end{equation}
Moreover, we neglect the contribution from the time range $t<t_{p}$,
where $t_{p}$ is the time when the scale $p$ exits the horizon since
$\zeta_{p}$ is oscillatory before the horizon exit. Using the super-horizon
approximation for mode functions during inflation
\begin{eqnarray}
\zeta_{k}(t) & = & \frac{1}{\sqrt{4\epsilon}M_{p}}\frac{H}{k^{\frac{3}{2}}}e^{i\frac{k}{aH}}(1-i\frac{k}{aH}),\\
u_{k}(t) & \approx & a^{-\frac{3}{2}}H^{-\frac{1}{2}}\left\{ \frac{2^{\nu-1}\Gamma(\nu)}{\pi^{\frac{1}{2}}}\left(\frac{k}{aH}\right)^{-\nu}+i\frac{\pi^{\frac{1}{2}}}{2^{\nu+1}\Gamma(1+\nu)}\left(\frac{k}{aH}\right)^{\nu}\right\} ,
\end{eqnarray}
where $\nu\equiv\sqrt{9/4-m^{2}/H^{2}}$, the cross-correlation at
the end of inflation time $t_{e}$ is approximately
\begin{equation}
\widetilde{\left\langle \left(\sigma^{2}\right)_{r}\zeta\right\rangle _{p}^{C}}\sim\frac{-1}{8\pi^{2}}\left|\zeta_{p}^{o}\right|^{2}\frac{H^{4}}{m_{\sigma}^{2}}\left[1-\left(\frac{p}{a_{e}H}\right)^{\frac{2m_{\sigma}^{2}}{3H^{2}}}\right]\label{eq:naive_sigma2zeta}
\end{equation}
where we used the relations $m_{\sigma}^{2}\ll H^{2}$ and $\left|\zeta_{p}^{o}\right|^{2}p^{3}=H^{2}/4M_{p}^{2}\epsilon$
is the mode function behavior in the long wavelength limit. To understand
the magnitude of this expression, note that for physical CMB scale
comoving momenta, we have 
\begin{equation}
\frac{p}{a_{e}}=e^{-N(p)}H
\end{equation}
 for $N(p)\sim O(50)$. As long as 
\begin{equation}
1\gg m_{\sigma}^{2}/H^{2}\gtrsim1/N(p),\label{eq:msigoverhbound}
\end{equation}
we can estimate
\begin{equation}
\widetilde{\left\langle \left(\sigma^{2}\right)_{r}\zeta\right\rangle _{p}^{C}}\sim\frac{-1}{8\pi^{2}}\left|\zeta_{p}^{o}\right|^{2}\frac{H^{4}}{m_{\sigma}^{2}}
\end{equation}
which is an expression that is valid when the $p$ is far outside
of the horizon and a constant $H$ is a good approximation. Note that
this does not vanish in the limit $p\rightarrow0$. We will soon see
that this non-vanishing behavior is incorrect and is a signal of explicit
breaking diffeomorphism invariance coming from Eq.~(\ref{eq:uvcutoff}).
Note that if Eq.~(\ref{eq:msigoverhbound}) is not satisfied because
$m_{\sigma}=0$, we have
\begin{eqnarray}
\widetilde{\left\langle \left(\sigma^{2}\right)_{r}\zeta\right\rangle _{p}^{C}} & \sim & \frac{H^{2}}{12\pi^{2}}\left|\zeta_{p}^{o}\right|^{2}\ln\frac{p}{a_{e}H}\label{eq:dummy2}\\
 & \sim & -N(p)\frac{H^{2}}{12\pi^{2}}\left|\zeta_{p}^{o}\right|^{2}\label{eq:masslesscrosscorrbad}
\end{eqnarray}
 which again does not vanish and is negative.

As explained around Eq.~(\ref{eq:physicalroleofbeta}), the importance
of the cross-correlation in the isocurvature bound depends on whether
$\beta$ is of order $10^{-2}$ or larger and not by whether the cross
correlation by itself is of the order of curvature perturbations.
To compute $\beta$ defined in Eq.~(\ref{eq:fractional_cross-correlation}),
we need an estimate of $(\sigma^{2})_{r}$ correlator which we can
take from \cite{Chung:2011xd}: 
\begin{equation}
\widetilde{\left\langle \left(\sigma^{2}\right)_{r}\left(\sigma^{2}\right)_{r}\right\rangle _{p}^{C}}\sim\frac{1}{2\pi^{2}}\frac{H^{4}}{p^{3}}f(m_{\sigma}/H,p/a_{e}H)\label{eq:sigcorrelator}
\end{equation}
where $f$ is a function which can have an exponentially small value
owing to the functional behavior
\begin{equation}
f\sim\frac{H^{2}}{m_{\sigma}^{2}}\left(\frac{p}{a_{e}H}\right)^{\frac{4}{3}\frac{m_{\sigma}^{2}}{H^{2}}}.\label{eq:fexplicit}
\end{equation}
Combining Eqs.~(\ref{eq:fractional_cross-correlation}), (\ref{eq:naive_sigma2zeta}),
and (\ref{eq:sigcorrelator}), we find
\begin{eqnarray}
\beta^{\mbox{wrong}} & \sim & \sqrt{\Delta_{\zeta}^{2}}\frac{H}{4m_{\sigma}}\left(\frac{p}{a_{e}H}\right)^{-\frac{2}{3}\frac{m_{\sigma}^{2}}{H^{2}}}\label{eq:dummy3}\\
 & \sim & \frac{H}{4m_{\sigma}}e^{\frac{2}{3}\frac{m_{\sigma}^{2}}{H^{2}}N-12}\label{eq:betaenhanced}
\end{eqnarray}
which after recalling that $N\sim O(50$) and Eq.~(\ref{eq:msigoverhbound})
gives some hope that a proper computation would give a large value
for $\beta$ with $m_{\sigma}/H$ satisfying Eq.~(\ref{eq:msigoverhbound}).\footnote{It is important to keep in mind that we are making an assumption here about the isocurvature evolution when identifying the primordial computations of Eqs.~(\ref{eq:naive_sigma2zeta}) and (\ref{eq:sigcorrelator}) with the CMB observables of Eq.~(\ref{eq:fractional_cross-correlation}) where $c_i$ are computed according to the simple transfer treatment of Appendix \ref{sec:Behaviors-of-Transfer}.  We will discuss this assumption more in detail in subsection \ref{sub:Weakness}.}
For example, if $|\beta|=O(1)$, then any appreciable isocurvature
perturbation would be ruled out with the current data, affecting predictions
of \cite{Hikage:2008sk,Kawasaki:2008sn,Chung:2011xd}.

Recall from Eq.~(\ref{eq:physicalroleofbeta}) that the role of the
cross correlation can become important if $\beta$ can become sizable
while keeping $\alpha$ also sizable. One may worry that the enhancement
factor in $\beta$ of Eq.~(\ref{eq:sigcorrelator}) which is approximately
proportional to $\alpha$ may make $\alpha$ negligible in the parameter
regime in which $\beta$ is enhanced. However, note that $\alpha$
is controlled not just by Eq.~(\ref{eq:sigcorrelator}) but by 
\begin{equation}
\widetilde{\left\langle \delta_{S}\delta_{S}\right\rangle _{p}}=\frac{\widetilde{\left\langle \left(\sigma^{2}\right)_{r}\left(\sigma^{2}\right)_{r}\right\rangle _{p}^{C}}}{\left[\left\langle \left(\sigma^{2}\right)_{r}\right\rangle \right]^{2}}\label{eq:emphasizedivision}
\end{equation}
which has a one point function squared in the denominator proportional
to the energy density squared of $\sigma$. One can straight forwardly
check from Ref.~\cite{Chung:2011xd} that the denominator of Eq.~(\ref{eq:emphasizedivision})
can be tuned such that $\alpha$ can remain constant while $\widetilde{\left\langle \left(\sigma^{2}\right)_{r}\left(\sigma^{2}\right)_{r}\right\rangle _{p}^{C}}$
is sufficiently small as to enhance $\beta$ as described in Eq.~(\ref{eq:betaenhanced}).

Given this generic possibility of ruling out a large class of isocurvature
perturbation models, we consider below the leading gravitational interaction
contribution to $\beta$ carefully. We find that unlike the naive
estimate given in Eq.~(\ref{eq:naive_sigma2zeta}), there is a suppression
in the limit $p/(aH)\rightarrow0$ for the mass in the range of Eq.~(\ref{eq:msigoverhbound}).
The suppression in the numerator of $\beta$ precisely cancels the
denominator suppression factor coming from $f$ in Eq.~(\ref{eq:fexplicit})
such that no enhancement is obtained, contrary to the naive expectation
of Eq.~(\ref{eq:betaenhanced}). This suppression of the numerator
in the proper computation not seen in the naive estimate can be attributed
to a Ward identity associated with the diffeomorphism group element
of constant scaling of the spatial coordinates. Furthermore, a careful
computation that we give below will show that the sign of the cross-correlation
will be opposite to the naive estimate, owing to the fact that the
cross correlation here is tied to particle production instead of volume
dilution.

The detailed computation will address also explicitly how same answer
to the gauge invariant correlator results in two different gauges
of comoving gauge and uniform curvature gauge (one can verify this
is not obvious from the naive estimate presented in this subsection).
Another technical care that is taken in the computations below is
to explicitly specify how diffeomorphism invariant counter terms are
introduced to renormalize the composite operators intrinsic to $\delta_{S}$.
Since the correct answer relies on a gravitational Ward identity,
identifying proper diffeomorphism invariant regulator and counter
terms is important for a trustworthy computation. On the other hand,
note that the finite parts of the counter terms that remain after
the divergences are canceled will not affect the results to the leading
$\hbar$ expansion that we are concerned with.%
\footnote{Note that particle production is non-perturbative in $\hbar$. %
}

\subsection{Plausible and Correct Estimation Using a Soft-$\zeta$ Theorem}

Before we describe the actual computation, we give in this subsection
a method akin to the soft-$\zeta$ theorem used by \cite{Maldacena:2002vr,Cheung:2007st,Li:2008gg,Creminelli:2011mw,Creminelli:2011rh,Maldacena:2011nz,Antoniadis:2011ib,Kehagias:2012pd,Creminelli:2012ed,Assassi:2012et,Assassi:2012zq,Sugiyama:2012tr,Senatore:2012wy,Kehagias:2013yd}
to estimate the correct answer without a detailed computation. We
will also point out what ad-hoc assumptions are needed to make this
estimate using this theorem. A rigorous computation will be given
in subsection (\ref{sub:Two-point-Correlators}).

In the soft-$\zeta$ theorem application to the correlators in inflation,
one factorizes $N$-point function including at least one soft external
$\zeta$ into $(N-1)$-point function times the two point function
$\left\langle \zeta\zeta\right\rangle $. The well-known example is
the three-point function $\left\langle \zeta\zeta\zeta\right\rangle $
in the squeezed limit in quasi-dS space:
\begin{equation}
\int\frac{d^{3}q}{(2\pi)^{3}}\left\langle \zeta_{\vec{q}}\zeta_{\vec{k}}\zeta_{\vec{p}}\right\rangle \overset{p\to0}{\longrightarrow}-\left|\zeta_{p}^{o}\right|^{2}\frac{1}{k^{3}}\frac{\partial}{\partial\ln k}\left[k^{3}\widetilde{\left\langle \zeta\zeta\right\rangle _{k}}\right]\sim-\left(n_{s}-1\right)\left|\zeta_{p}^{o}\right|^{2}\left|\zeta_{k}^{o}\right|^{2}\label{eq:maldacena}
\end{equation}
where the superscript on the $\zeta$ mode functions denote long wavelength
parts. To use this, note that if we neglect renormalization of the
composite operators, we can write
\begin{equation}
\int\frac{d^{3}q}{(2\pi)^{3}}\langle\zeta_{\vec{p}}\sigma^{2}(\vec{q})\rangle=\int\frac{d^{3}k_{2}}{(2\pi)^{3}}\int\frac{d^{3}k_{1}}{(2\pi)^{3}}\langle\zeta_{\vec{p}}\sigma(\vec{k}_{1})\sigma(\vec{k}_{2})\rangle.
\end{equation}
Using Eq.~(\ref{eq:maldacena}) and replacing two $\zeta$ fields
with $\sigma$ fields, we can estimate
\begin{equation}
\int\frac{d^{3}q}{(2\pi)^{3}}\langle\zeta_{\vec{p}}\sigma^{2}(\vec{q})\rangle\overset{p\to0}{\longrightarrow}-\left|\zeta_{p}^{o}\right|^{2}\int_{p}\frac{d^{3}k_{2}}{(2\pi)^{3}}\frac{1}{k_{2}^{3}}\frac{\partial}{\partial\ln k_{2}}\left[k_{2}^{3}\widetilde{\left\langle \sigma\sigma\right\rangle _{k_{2}}}\right]\label{eq:usingmaldacena}
\end{equation}
where the comoving IR cutoff $p$ is required to treat $\zeta_{p}^{o}$
as a constant background field. This effective lower cutoff $p$ cannot
be justified without explicit computation, but this is physically
plausible because $\langle\sigma\sigma\rangle$ does not have any
IR divergence as long as $m_{\sigma}^{2}>0$. One can rewrite the
integral in Eq.~(\ref{eq:usingmaldacena}) as 
\begin{equation}
\int\frac{d^{3}q}{(2\pi)^{3}}\left\langle \zeta_{\vec{p}}\sigma^{2}(\vec{q})\right\rangle \overset{p\to0}{\longrightarrow}\left|\zeta_{p}^{o}\right|^{2}\frac{\partial}{\partial\ln a}\left\langle \sigma^{2}(t,\vec{x})\right\rangle _{p}\label{eq:soft-z_arg-2pt_fn}
\end{equation}
where the $\sigma^{2}$ on the right hand side corresponds to spacetime
field (and not its Fourier transform), the $p$ subscript on the bracket
corresponds to the IR cutoff in the mode function integral, and we
assume that there is no contribution from the UV cutoff. It is easy
to prove that if $p\rightarrow0$ is well defined and a UV cutoff
is not required, then the right hand side of Eq.~(\ref{eq:soft-z_arg-2pt_fn})
vanishes in the limit $p\rightarrow0$. This is in contrast with Eq.~(\ref{eq:naive_sigma2zeta}). 

The vanishing of this function in the $p\rightarrow0$ limit for $m_{\sigma}^{2}>0$
is intuitively understood from the fact that in that limit, $\zeta_{p}^{o}$
acts as a spatial diffeomorphism
\begin{equation}
\vec{x}\rightarrow\vec{x}(1+\zeta_{p}^{0})\label{eq:diffeo}
\end{equation}
(which in turn effectively rescales the scale factor $a$ by a constant
factor if we neglect spatial derivatives on long wavelengths) which
cannot change $\left\langle \sigma^{2}(t,\vec{x})\right\rangle =\left\langle \sigma^{2}(t,0)\right\rangle .$
More explicitly, one can show that the explicit computation can be
rewritten as
\begin{equation}
\int\frac{d^{3}q}{(2\pi)^{3}}\left\langle \zeta_{\vec{p}}\sigma^{2}(\vec{q})\right\rangle \overset{p\to0}{\longrightarrow}|\zeta_{p}|^{2}\int_{p}\frac{d^{3}k}{(2\pi)^{3}}\int d^{3}xi\langle[\hat{Q}(t),\hat{\sigma}(t,\vec{x})\hat{\sigma}(t,0)]\rangle e^{i\vec{k}\cdot\vec{x}}
\end{equation}
where 
\begin{equation}
\hat{Q}(t)\equiv\int^{t}d^{4}za^{2}(t_{z})\delta_{ij}\hat{T}_{\sigma}^{ij}(z)
\end{equation}
is the generator of the diffeomorphism associated with Eq.~(\ref{eq:diffeo}).
Note that the right hand side formally vanishes when the IR cutoff
is removed (i.e. $p=0$) because in that limit, we find the commutator
\begin{equation}
\langle[\hat{Q}(t),\hat{\sigma}^{2}(t,0)]\rangle=0.\label{eq:diffeo-1}
\end{equation}
This can be interpreted also as a Ward identity. On the flip side,
as long as $p\neq0$, $\left\langle \sigma^{2}(t,\vec{x})\right\rangle _{p}$
is not invariant under the diffeomorphism Eq.~(\ref{eq:diffeo}).
The crucial point from this perspective is that diffeomorphism invariance
is extremely important to see that the cross correlation vanishes
for $p\rightarrow0$ for a massive scalar field. It is this that one
failed to preserve in Eq.~(\ref{eq:uvcutoff}).

As we will show in detail, Eq.~(\ref{eq:soft-z_arg-2pt_fn}) is consistent
with the explicit computation. Note that a couple of assumptions that
we already mentioned in deriving Eq.~(\ref{eq:soft-z_arg-2pt_fn})
can only be justified by an explicit computation: namely, the effective
lower cutoff $p$ in Eq.~(\ref{eq:usingmaldacena}) and UV cutoff
details associated with renormalizing the composite operator $\sigma^{2}$.
Such complications do not arise in isocurvature scenarios without
composite operators. Hence, one of the main technical merits of this
paper is to provide a explicit justification of Eq.~(\ref{eq:soft-z_arg-2pt_fn}).
Note that because the diffeomorphism gauge invariance plays a crucial
role in obtaining the correct $p$ dependence in Eq.~(\ref{eq:soft-z_arg-2pt_fn})
as explained around Eq.~(\ref{eq:diffeo-1}), we choose a UV regulator
that preserves diffeomorphism invariance in the computation below.

\subsection{Gauge Invariance of Correlators}

\label{sec:gauge-inv}Before we begin our explicit computation, we
will check the setup of our computation by demonstrating that the
manifestly gauge invariant quantities $\langle\delta_{S}\zeta\rangle$
and $\langle\delta_{S}\delta_{S}\rangle$ yield the same values in
comoving and in the uniform curvature gauges. To accomplish this,
we use a gravitational Ward identity.

We first note that the $\zeta$ dependent metric perturbations $\delta g^{(C)}$
and $\delta g^{(U)}$ differs by a gauge transformation, i.e. 
\begin{equation}
\Delta g_{\mu\nu}=\delta g_{\mu\nu}^{(U)}-\delta g_{\mu\nu}^{(C)}=\left(\begin{array}{cc}
2\frac{d}{dt}(\frac{\zeta}{H}) & (-\frac{\zeta}{H})_{,i}\\
(-\frac{\zeta}{H})_{,i} & -a^{2}\delta_{ij}2\zeta
\end{array}\right)=-[\mathcal{L}_{X}\bar{g}]_{\mu\nu},
\end{equation}
where 
\begin{equation}
X^{0}=-\frac{\zeta}{H},\quad X^{i}=0.
\end{equation}
Their interaction actions differ by 
\begin{eqnarray}
\Delta S_{\sigma\sigma\zeta}=S_{\sigma\sigma\zeta}^{(U)}-S_{\sigma\sigma\zeta}^{(C)} & = & -\int^{t_{f}}dtd^{3}xa_{x}^{3}\, T^{\mu\nu}(\bar{g},\sigma)\nabla_{\mu}X_{\nu}
\end{eqnarray}
Their interaction Hamiltonians differ by 
\begin{eqnarray}
\Delta H_{\zeta\sigma\sigma}(t)=H_{\zeta\sigma\sigma}^{(U)}(t)-H_{\zeta\sigma\sigma}^{(C)}(t) & = & \int d^{3}x\, a^{3}(t)T^{\mu\nu}(\bar{g},\sigma;t,\vec{x})\nabla_{\mu}X_{\nu}(t,\vec{x})
\end{eqnarray}

Then we compare $\langle\sigma_{x}^{2}\zeta_{y}\rangle$ in the two
gauges:
\begin{eqnarray}
\langle\sigma^{2}(t_{f},\vec{x})\zeta(t_{f},\vec{y})\rangle^{U}-\langle\sigma^{2}(t_{f},\vec{x})\zeta(t_{f},\vec{y})\rangle^{C} & = & -i\int^{t_{f}}dt\left\langle \left[\sigma_{x}^{2}\zeta_{y},\Delta H_{\zeta\sigma\sigma}(t)\right]\right\rangle \label{eq:dummy6}\\
 & = & -i\int^{t_{f}}dtd^{3}z\left\langle \left[\sigma_{x}^{2}\zeta_{y},\nabla_{\mu}\left(a^{3}(t)T^{\mu\nu}(\bar{g},\sigma;t,\vec{x})X_{\nu}(t,\vec{x})\right)\right]\right\rangle \label{eq:commutatorintermed}
\end{eqnarray}
where we have integrated by parts and used the quantum version of
$\nabla_{\mu}T_{\sigma}^{\mu\nu}=0$: i.e. in-in formalism gravitational
Ward identities
\begin{eqnarray}
i\nabla_{\mu}\langle in|T_{z}^{\mu\nu+}\sigma_{x}^{+}\sigma_{y}^{+}|in\rangle_{g} & = & \frac{1}{\sqrt{g_{x}}}\delta^{4}(x-z)g_{x}^{\alpha\nu}\frac{\partial}{\partial x^{\alpha}}\langle in|\sigma_{x}^{+}\sigma_{y}^{+}|in\rangle_{g}\nonumber \\
 &  & +\frac{1}{\sqrt{g_{y}}}\delta^{4}(y-z)g_{y}^{\alpha\nu}\frac{\partial}{\partial y^{\alpha}}\langle in|\sigma_{x}^{+}\sigma_{y}^{+}|in\rangle_{g}\label{eq:dummy7}\\
i\nabla_{\mu}\langle in|T_{z}^{\mu\nu-}\sigma_{x}^{+}\sigma_{y}^{+}|in\rangle_{g} & = & 0\label{eq:dummy8}
\end{eqnarray}
whose the notation is explained in Section \ref{sec:Review-of-Diffeomorphism}.
Note that the remaining term in Eq.~(\ref{eq:commutatorintermed})
is a total derivative. Hence, we are left with the boundary contribution
\begin{eqnarray}
\langle\sigma^{2}(t_{f},\vec{x})\zeta(t_{f},\vec{y})\rangle^{U}-\langle\sigma^{2}(t_{f},\vec{x})\zeta(t_{f},\vec{y})\rangle^{C} & = & -i\int d^{3}z\, a^{3}(t_{f})\frac{1}{H}\left\langle \left[\sigma_{x}^{2},T_{\sigma,z}^{00}\right]\right\rangle \langle\zeta_{z}\zeta_{y}\rangle\label{eq:dummy9}\\
 & = & -\frac{\partial_{t}\langle\sigma_{x}^{2}\rangle}{H}\langle\zeta_{x}\zeta_{y}\rangle.\label{eq:Gauge_Inv}
\end{eqnarray}
To make these composite operator correlators well defined while maintaining
diffeomorphism invariance (see the discussion surrounding Eq.~(\ref{eq:diffeo-1})),
we need a proper covariant regulator, such as the Pauli-Villars (PV)
regulator. It is straightforward to use the PV regulator here because
the above identity holds for PV fields as well. See Appendix \ref{sec:Renormalization-of-Composite-Operators}
for a more detailed discussion of the prescription of the PV regulator. 

Using Eq.~(\ref{eq:Gauge_Inv}), it is now trivial to show that $\langle\delta_{S}\zeta\rangle^{U}=\langle\delta_{S}\zeta\rangle^{C}$
and $\langle\delta_{S}\delta_{S}\rangle^{U}=\langle\delta_{S}\delta_{S}\rangle^{C}$.
Because $\delta_{S}\ni\sigma_{x}^{2}/\langle\sigma_{x}^{2}\rangle$,
the denominator of this expression also transforms:
\begin{equation}
\Delta\delta_{S}\ni-\frac{\Delta\langle\sigma_{x}^{2}\rangle}{\langle\sigma_{x}^{2}\rangle}\frac{\sigma_{x}^{2}}{\langle\sigma_{x}^{2}\rangle}=\frac{\zeta_{x}}{H}\frac{\partial_{t}\langle\sigma_{x}^{2}\rangle}{\langle\sigma_{x}^{2}\rangle}\frac{\sigma_{x}^{2}}{\langle\sigma_{x}^{2}\rangle}\label{eq:dummy10}
\end{equation}
which leads to a cancellation of Eq.~(\ref{eq:Gauge_Inv}) consistently
to leading $\hbar\rightarrow0$ approximation. Hence, we have a nontrivial
consistency check of the computation that we are setting up.

\subsection{Two-point Functions\label{sub:Two-point-Correlators}}

In this subsection, we present a rigorous computation of $\beta$
defined in (\ref{eq:fractional_cross-correlation}). To this end,
we need to calculate the two-point function $\left\langle \left(\sigma^{2}\right)_{r}\zeta\right\rangle $
and $\left\langle \left(\sigma^{2}\right)_{r}\left(\sigma^{2}\right)_{r}\right\rangle $
where the renormalized composite operator \cite{DeWitt:1975ys,birrell1982ix,Fulling:1989nb,Horowitz:1980fk,Liddle:2000cg,Hu:2003qn,Finelli:2007fr,Prokopec:2008gw,PerezNadal:2009hr,Baacke:2010bm,Ford:2010wd,Agullo:2011qg,Wu:2011gk}
is 
\begin{equation}
(\sigma^{2})_{r}\equiv(\sigma+\sum_{n}\chi_{n})^{2}+\delta Z_{0}(\Lambda,m_{\sigma})+\delta Z_{1}(\Lambda,m_{\sigma})R\label{eq:dummy11}
\end{equation}
which is discussed in greater detail in Sec.~(\ref{sec:ren-com-op}).
Here we are going to use the comoving gauge for the computation because
of its advantages that we state below.%
\footnote{This computation has been done also in the uniform curvature gauge,
which is presented in Appendix \ref{sec:Two-Point-Functions-in-UG}.
Particularly, in the massless limit, we explicitly calculate up to
the next leading term including all gravitational couplings. This
shows that the next leading terms are indeed suppressed by the factor
$p^{2}/a^{2}$.%
} As shown in Eqs. (\ref{eq:Hint}) and (\ref{eq:deltag_CG}), the
gravitational interactions in the comoving gauge are derivatively
(i.e. $p^{2}/a^{2}$) suppressed except the $(ij)$-components. In
other words, the contributions from $T_{\sigma}^{00}\delta g_{00}^{(C)}$
and $T_{\sigma}^{0i}\delta g_{0i}^{(C)}$ interactions are $O(p^{2}/a^{2}),$
where $\vec{p}$ is an external 3-momentum. Furthermore, all counter
term contributions are also derivatively suppressed in the comoving
gauge: $\delta Z_{0}\left\langle \zeta\right\rangle =0$ and $\delta Z_{1}\widetilde{\left\langle R\zeta\right\rangle _{p}^{C}}=O(p^{2}/a^{2})$.
Therefore, we don't need the counter terms to compute the non-derivatively
suppressed contributions, but we still need a regulator for UV divergences
in the computation. The regulator dependences and the UV divergences
will automatically disappear together in our final result.

\begin{figure}
\begin{centering}
\includegraphics{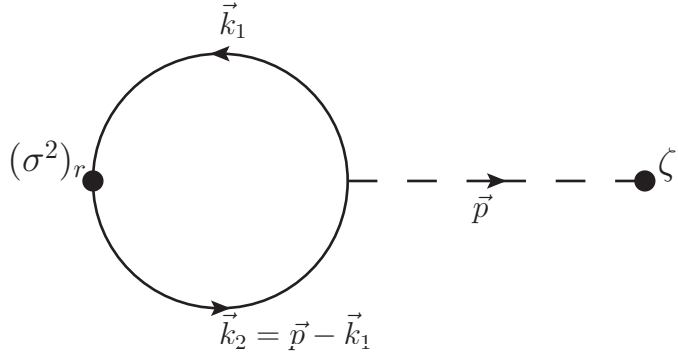}
\par\end{centering}

\caption{\label{fig:Two-point-function}Two-point function at one loop order. }
\end{figure}

Now we compute the two-point function shown in Fig. \ref{fig:Two-point-function},
which is written in the Fourier space as
\begin{eqnarray}
\widetilde{\left\langle \left(\sigma^{2}\right)_{r}\zeta\right\rangle _{p}^{C}} & = & \int d^{3}x\, e^{-i\vec{p}\cdot\vec{x}}\left\langle \left(\sigma^{2}(t,\vec{x})\right)_{r}\zeta(t,\vec{0})\right\rangle ^{C}\label{eq:dummy12}\\
 & = & \int d^{3}x\, e^{-i\vec{p}\cdot\vec{x}}\int^{t}d^{4}z\, a^{3}(t_{z})\sum_{N=0}^{s}\left\langle \left[\sigma_{N}^{2}(t,\vec{x})\zeta(t,\vec{0}),\frac{i}{2}\left(2\zeta a^{2}\delta_{ij}T_{\sigma}^{ij}\right)_{z}\right]\right\rangle +O\left(\frac{p^{2}}{a^{2}}\right),
\end{eqnarray}
where we have introduced the Pauli-Villars (PV) regulator (see Appendix
\ref{sec:Renormalization-of-Composite-Operators} for more details)
and 
\begin{eqnarray}
a^{2}\delta_{ij}T_{\sigma}^{ij} & = & -3\mathcal{L}_{\sigma}+\sum_{N=0}^{s}C_{N}\left(\frac{\nabla}{a}\sigma_{N}\right)^{2},\label{eq:Tij}
\end{eqnarray}
where $\sigma_{0}$ and $\sigma_{n}$ are the physical field $\sigma$
and the PV field $\chi_{n}$ (here, $n\in\{1,2,...,s\}$), respectively,
and $s$ is the number of introduced PV fields.

Interestingly, this integral can be computed in any FRW space-time.
We first compute the second term contribution in Eq.~(\ref{eq:Tij})
defined as
\begin{equation}
I_{N}^{(2)}(p)\equiv\int d^{3}x\, e^{-i\vec{p}\cdot\vec{x}}\int^{t}d^{4}z\sqrt{-g_{z}}\left\langle \left[\sigma_{N}^{2}(t,\vec{x})\zeta(t,\vec{0}),i\zeta_{z}C_{N}\left(\frac{\nabla}{a}\sigma_{N}\right)_{z}^{2}\right]\right\rangle \label{eq:dummy13}
\end{equation}
 Expanding in mode functions, this becomes
\begin{eqnarray}
I_{N}^{(2)}(p) & = & -4C_{N}^{-1}\int\frac{d^{3}k_{1}}{\left(2\pi\right)^{3}}d^{3}k_{2}\delta^{3}(\vec{k}_{1}+\vec{k}_{2}-\vec{p})\int_{-\infty}^{t}dt_{z}a_{z}^{3}\nonumber \\
 &  & \times\left(-\frac{\vec{k}_{1}\cdot\vec{k}_{2}}{a_{z}^{2}}\right)\mbox{Im}\left[\zeta_{p}(t)\zeta_{p}^{*}(t_{z})u_{N,k_{1}}(t)u_{N,k_{1}}^{*}(t_{z})u_{N,k_{2}}(t)u_{N,k_{2}}^{*}(t_{z})\right],
\end{eqnarray}
where $u_{N}$ are the mode functions for fields $\sigma_{N}$. Because
$\zeta$ oscillates before and freezes after the horizon exit, we
neglect the contribution before the horizon exit. Furthermore, we
can neglect the $O(p^{2}/a^{2})$ term and factor $\zeta_{p}$ out
of the time integral. We thus find
\begin{eqnarray}
I^{(2)}(p) & \approx & 4\left|\zeta_{p}^{o}(t)\right|^{2}\int\frac{d^{3}k_{1}}{\left(2\pi\right)^{3}}d^{3}k_{2}\delta^{3}(\vec{k}_{1}+\vec{k}_{2}-\vec{p})\nonumber \\
 &  & \times\int_{t_{p}}^{t}dt_{z}a_{z}^{3}\left(\frac{\vec{k}_{1}\cdot\vec{k}_{2}}{a_{z}^{2}}\right)\mbox{Im}\left[u_{k_{1}}(t)u_{k_{1}}^{*}(t_{z})u_{k_{2}}(t)u_{k_{2}}^{*}(t_{z})\right]+O\left(\frac{p^{2}}{a^{2}}\right),\label{eq:dummy14}
\end{eqnarray}
where $t_{p}$ is the time at which scale $p$ exits the horizon.
Note that we drop subscript $N$ and field normalization $C_{N}$
for convenience, but we will put it back later in the final result.
Moreover, we neglect the low momentum phase space, i.e. $\mbox{\ensuremath{min}\{}k_{1},k_{2}\}<p$,
because of $\left|u_{k}\right|^{2}\lesssim O(k^{-3})$ and the spatial
gradient factor $\vec{k}_{1}\cdot\vec{k}_{2}/a^{2}$. 
\begin{equation}
\int_{k_{1}<p}\frac{d^{3}k_{1}}{\left(2\pi\right)^{3}}d^{3}k_{2}\delta^{3}(\vec{k}_{1}+\vec{k}_{2}-\vec{p})\int_{t_{p}}^{t}dt_{z}a_{z}^{3}\left(\frac{\vec{k}_{1}\cdot\vec{k}_{2}}{a_{z}^{2}}\right)\mbox{Im}\left[u_{k_{1}}(t)u_{k_{1}}^{*}(t_{z})u_{k_{2}}(t)u_{k_{2}}^{*}(t_{z})\right]\lesssim O\left(\frac{p^{2}}{a^{2}}\right).
\end{equation}
Then the main contribution of the integral comes from the phase space
$k_{1},k_{2}>p$, and thus $p$ behaves as an IR cut-off (see the
importance of this IR cutoff in the discussion surrounding Eq.~(\ref{eq:soft-z_arg-2pt_fn})). 

Since $k_{1},k_{2}>p$, we Taylor-expand the integrand with respect
to $p$ and take the leading term. Then we have
\begin{equation}
I^{(2)}(p)\approx4\left|\zeta_{p}^{o}(t)\right|^{2}\int_{p}\frac{d^{3}k_{1}}{\left(2\pi\right)^{3}}\int_{t_{p}}^{t}dt_{z}a_{z}^{3}\left(-\frac{k_{1}^{2}}{a_{z}^{2}}\right)\mbox{Im}\left[u_{k_{1}}^{2}(t)u_{k_{1}}^{*2}(t_{z})\right]+O\left(\frac{p^{2}}{a^{2}}\right).\label{eq:Int_partial_sigma2}
\end{equation}
Now we are going to compute the time integral. Recall that the differential
equation for mode function $u_{k}$ is
\begin{equation}
\ddot{u}_{k}+3Hu_{k}+\left(\frac{k^{2}}{a^{2}}+m^{2}\right)u_{k}=0.
\end{equation}
Applying $\frac{\partial}{\partial\ln k}$ to the equation, we obtain
\begin{equation}
\ddot{y}_{k}+3Hy_{k}+\left(\frac{k^{2}}{a^{2}}+m^{2}\right)y_{k}=-2\frac{k^{2}}{a^{2}}u_{k},
\end{equation}
where $y_{k}\equiv\frac{\partial}{\partial\ln k}u_{k}.$ Note that
the homogeneous solutions for $y_{k}$ are $u_{k}$ and $u_{k}^{*}$.
Thus, we use the Green function method to find a solution 
\begin{equation}
y_{k}(t)=\int^{t}dt'\frac{a^{3}(t')}{i}\left(u_{k}^{*}(t)u_{k}(t')-u_{k}(t)u_{k}^{*}(t')\right)\left(-2\frac{k^{2}}{a^{2}}\right)u_{k}(t').
\end{equation}
From this, we find 
\begin{eqnarray}
\frac{d}{d\ln k}\left|u_{k}(t)\right|^{2} & = & 2\mbox{Re}\left[u_{k}^{*}(t)y_{k}(t)\right]\label{eq:dummy15}\\
 & = & 4\int_{-\infty}^{t}dt_{z}a_{z}^{3}\frac{k^{2}}{a_{z}^{2}}\mbox{Im}\left[u_{k}^{2}(t)u_{k}^{*2}(t_{z})\right]\label{eq:dummy15.5}\\
 & = & \left[\int_{t_{p}}^{t}dt_{z}+\int_{-\infty}^{t_{p}}dt_{z}\right]4a_{z}^{3}\frac{k^{2}}{a_{z}^{2}}\mbox{Im}\left[u_{k}^{2}(t)u_{k}^{*2}(t_{z})\right].
\end{eqnarray}
The second term is oscillatory with respect to $k$ so that we can
safely neglect it after the momentum integral. Inserting this back
to the integral (\ref{eq:Int_partial_sigma2}), we obtain
\begin{eqnarray}
I_{N}^{(2)}(p) & \approx & -C_{N}^{-1}\left|\zeta_{p}^{o}(t)\right|^{2}\int_{p}\frac{d^{3}k}{\left(2\pi\right)^{3}}\frac{d}{d\ln k}\left|u_{N,k}(t)\right|^{2}+O\left(\frac{p^{2}}{a^{2}}\right)\label{eq:dummy16}\\
 & = & -C_{N}^{-1}\left|\zeta_{p}^{o}(t)\right|^{2}\left[-\left.\frac{k^{3}}{2\pi^{2}}\left|u_{N,k}(t)\right|^{2}\right|_{p}^{\Lambda_{UV}}+3\left\langle \left(\sigma_{N}^{2}\right)_{p}\right\rangle \right]+O\left(\frac{p^{2}}{a^{2}}\right),\label{eq:Res_Int_2}
\end{eqnarray}
where we have put the subscript $N$ and the field normalization $C_{N}$
back, and
\begin{equation}
\left\langle \left(\sigma_{N}^{2}\right)_{p}\right\rangle \equiv\int_{p}\frac{d^{3}k}{\left(2\pi\right)^{3}}\left|u_{N,k}(t)\right|^{2},
\end{equation}
where the subscript $p$ stands for the comoving IR cut-off of momentum.
One can then compute the contribution of the first term in Eq. (\ref{eq:Tij})
in a similar manner:

\begin{eqnarray}
I_{N}^{(1)} & \equiv & \int d^{3}xe^{-i\vec{p}\cdot\vec{x}}\int^{t}d^{4}z\sqrt{-g_{z}}\left\langle \left[\sigma_{N}^{2}(t,\vec{x})\zeta(t,\vec{0}),i\left(-3\right)\mathcal{L}_{\sigma}(z)\zeta(z)\right]\right\rangle \label{eq:dummy17}\\
 & = & 3C_{N}^{-1}\left|\zeta_{p}^{o}\right|^{2}\left\langle \left(\sigma_{N}^{2}\right)_{p}\right\rangle +O\left(\frac{p^{2}}{a^{2}}\right).\label{eq:Res_Int_1}
\end{eqnarray}
Hence, we obtain
\begin{eqnarray}
\widetilde{\left\langle \left(\sigma^{2}\right)_{r}\zeta\right\rangle _{p}^{C}} & = & \sum_{N=0}^{s}I_{N}^{(1)}+I_{N}^{(2)}+O\left(\frac{p^{2}}{a^{2}}\right)\label{eq:dummy18}\\
 & = & \left|\zeta_{p}^{o}\right|^{2}\frac{p^{3}}{2\pi^{2}}\left|u_{p}(t)\right|^{2}+O\left(\frac{p^{2}}{a^{2}}\right)\label{eq:sigma2zeta_CG}
\end{eqnarray}
where $u_{p}$ is the mode function for physical field $\sigma$.

Comparing the computation of Eq.~(\ref{eq:sigma2zeta_CG}) with the
estimate in Sec.~\ref{sub:Puzzles}, we see two crucial differences:
\begin{enumerate}
\item There is a cancellation of the $3C_{N}^{-1}\left|\zeta_{p}^{o}\right|^{2}\left\langle \left(\sigma_{N}^{2}\right)_{p}\right\rangle $
term that is sensitive to mode summation that extends to sub horizon
modes.
\item The $\Lambda_{UV}$ dependent term in Eq.~(\ref{eq:Res_Int_2}) in
the present computation disappears after accounting for the PV regulator
fields. In contrast, the estimate in Sec.~\ref{sub:Puzzles} leaves
behind a $\Lambda_{UV}=aH_{inf}$ dependent contribution due to the
ad hoc nature of the UV cutoff which does not preserve diffeomorphism.
\end{enumerate}
Finally, putting the results (\ref{eq:Res_Int_2}) and (\ref{eq:Res_Int_1})
together, the two-point function becomes
\begin{eqnarray}
\left.\widetilde{\left\langle \left(\sigma^{2}\right)_{r}\zeta\right\rangle _{p}^{C}}\right|_{t_{e}} & = & \left|\zeta_{p}^{o}\right|^{2}\times\begin{cases}
\frac{\Gamma^{2}(\nu)H^{2}}{\pi^{3}}\left(\frac{p}{2a(t_{e})H}\right)^{3-2\nu} & \mbox{massive scalar in dS}\\
\frac{H_{p}^{2}}{4\pi^{2}} & \mbox{massless during quasi-dS}
\end{cases}+O\left(\frac{p^{2}}{a^{2}}\right)\label{eq:sigma2zeta_ds_CG}
\end{eqnarray}
where $H_{p}$ denote the Hubble scale at which scale $p$ exits the
horizon, $\nu=\sqrt{9/4-m^{2}/H^{2}}$, and $t_{e}$ reminds us that
we are evaluating this at the end of inflation. We have applied (quasi)-dS
mode function in evaluating (\ref{eq:sigma2zeta_ds_CG}).%
\footnote{\label{fn:After-inflation-ends}After inflation ends at time $t_{e}$,
the cross correlation is expressed as
\[
\widetilde{\left\langle \left(\sigma^{2}\right)_{r}\zeta\right\rangle _{p}^{C}}=f_{T}\left.\widetilde{\left\langle \left(\sigma^{2}\right)_{r}\zeta\right\rangle _{p}^{C}}\right|_{t_{e}}
\]
where $f_{T}$ accounts for the change in the mode-function behavior
after the end of inflation. As alluded to in the discussion near Eq.~(\ref{eq:physicalroleofbeta}),
the factor $f_{T}$ cancels out of the expression in $\beta$ due
to its appearance in the denominator $\sqrt{\Delta_{\zeta}^{2}\Delta_{\delta_{S}}^{2}}$.
The factor $f_{T}$ can also account for the corrections in the superhorizon
mode function behavior during inflation due to deviations away from
the exact dS background.%
} One can easily check that Eq.~(\ref{eq:naive_sigma2zeta}) is consistent
with Eq.~(\ref{eq:usingmaldacena}).

As explained near Eq.~(\ref{eq:diffeo-1}), the vanishing of the
cross-correlation in the limit $p\rightarrow0$ is expected from the
diffeomorphism Ward identity. For a nonvanishing $p$, one might expect
the cross-correlation should be $O(p^{2}/a^{2})$ by Taylor-expanding
the cross-correlation at $p=0$. However, Eq.~(\ref{eq:sigma2zeta_ds_CG})
interestingly shows that the leading term of the cross-correlation
is not analytic at $p=0$ and thus not $p^{2}/a^{2}$-suppresed. Indeed,
for any small $p/a(t_{e})$, we can diminish the suppression by making
$3-2\nu\rightarrow0^{+}$ through the limit $m/H\rightarrow0$. 

To finish the computation of $\beta$, we also consider the two-point
correlator $\left\langle \left(\sigma^{2}\right)_{r}\left(\sigma^{2}\right)_{r}\right\rangle $
showing up in the denominator. Again, the comoving gauge is convenient
for this computation. Although the correlator is UV divergent, because
the counter terms associated with the divergence are derivatively
suppressed, we do not need to include the counter terms in computing
the IR contributions and the non-derivative contribution of the correlator
is insensitive to renormalization. Furthermore, the IR contribution
using the super-horizon approximation is not UV divergent. That means
the UV contribution and the IR contribution are cleanly separated.
Thus, we can estimate $\widetilde{\left\langle \left(\sigma^{2}\right)_{r}\left(\sigma^{2}\right)_{r}\right\rangle }$
using only the super-horizon approximation unlike in the computation
of $\widetilde{\left\langle \left(\sigma^{2}\right)_{r}\zeta\right\rangle }$.
We find
\begin{equation}
\widetilde{\left\langle \left(\sigma^{2}\right)_{r}\left(\sigma^{2}\right)_{r}\right\rangle _{p}^{C}}=2\int_{\Lambda_{IR}}\frac{d^{3}k_{1}}{\left(2\pi\right)^{3}}d^{3}k_{2}\delta^{3}(\vec{k}_{1}+\vec{k}_{2}-\vec{p})\left|u_{k_{1}}(t)\right|^{2}\left|u_{k_{2}}(t)\right|^{2}+O\left(\frac{p^{2}}{a^{2}}\right)\label{eq:firsttimeircutoff}
\end{equation}
where $\Lambda_{IR}$ is a comoving IR cutoff. Evaluating this with
dS super horizon modes and assuming $m<3H/2$, we find the value at
the end of inflation to be

\begin{eqnarray}
\left.\widetilde{\left\langle \left(\sigma^{2}\right)_{r}\left(\sigma^{2}\right)_{r}\right\rangle _{p}^{C}}\right|_{t_{e}} & \approx & 2\int_{\Lambda_{IR}}\frac{d^{3}k_{1}}{\left(2\pi\right)^{3}}d^{3}k_{2}\delta^{3}(\vec{k}_{1}+\vec{k}_{2}-\vec{p})\frac{2^{-4+4\nu}\left|\Gamma(\nu)\right|^{4}}{\pi^{2}}\frac{1}{a^{6}(t_{e})H^{2}}\left(\frac{k_{1}}{a(t_{e})H}\right)^{-2\nu}\left(\frac{k_{2}}{a(t_{e})H}\right)^{-2\nu}\label{eq:dummy19}\\
 & \approx & \frac{1}{2\pi^{2}}\frac{H^{4}}{p^{3}}\frac{1}{3-2\nu}\left(\frac{p}{a(t_{e})H}\right)^{6-4\nu}\left[1-\left(\frac{\Lambda_{IR}}{p}\right)^{3-2\nu}\right].\label{eq:sigma2sigma2}
\end{eqnarray}

In Eq.~(\ref{eq:firsttimeircutoff}), we have introduced a comoving
IR cutoff $\Lambda_{IR}$ which corresponds to the statement that
inflationary era had a beginning in the finite past. Explicitly, we
cannot use the Bunch-Davies vacuum boundary condition for modes that
left the horizon before the beginning of inflation. This means that
\begin{equation}
\frac{\Lambda_{IR}}{p}\sim e^{-(N_{\mbox{tot}}-N(p))}\label{eq:dummy20}
\end{equation}
where $N_{\mbox{tot}}$ is the total number of efolds of inflation,
$N(p)$ is the number of efolds before the end of inflation at which
the mode $p$ left the horizon: i.e. $p/a(N)=H$. This cutoff is related
to the box cutoff introduced in \cite{Lyth:1991ub,Lyth:2007jh,Hikage:2008sk}.
Numerically, $\Lambda_{IR}\ll p$ is irrelevant when 
\begin{equation}
\frac{m_{\sigma}^{2}}{H^{2}}\gg\frac{1}{N_{\mbox{tot}}-N(p)}.\label{eq:IRcutoffinsensitivity}
\end{equation}
For situations in which this condition is violated, IR effects are
important, and our computation is only qualitatively suggestive since
$\Lambda_{IR}$ has to be resolved using more detailed description
of the beginning of inflation. In particular, since we do not physically
expect $N_{\mbox{tot}}=\infty$, $m_{\sigma}=0$ situation is not
accurately captured by our computation. Of course, the IR sensitivity
here is not important as far as the importance of the cross correlation
is concerned since the qualitative behavior of having $p/\Lambda_{IR}\rightarrow\infty$
is to make the correlation even larger making the $\beta$ parameter
even smaller. Finally, note that Eq.~(\ref{eq:IRcutoffinsensitivity})
can easily be more stringent than Eq.~(\ref{eq:msigoverhbound}).

Hence, we conclude
\begin{equation}
\beta\approx\begin{cases}
-\frac{m_{\sigma}}{H}\sqrt{\frac{\Delta_{\zeta}^{2}}{6}} & \mbox{massive scalar in dS}\\
-\frac{\sqrt{\Delta_{\zeta}^{2}}}{2}\left(\ln\frac{p}{\Lambda_{IR}}\right)^{-1/2} & \mbox{massless during quasi-dS}
\end{cases}\label{eq:betafinal}
\end{equation}
where for the massive scalar case is assume to satisfy Eq.~(\ref{eq:msigoverhbound}).
Although this in principle is a generic prediction of isocurvature
scenario, the magnitude of around $10^{-5}$ is difficult to probe
experimentally since the current sensitivity is at the level of $10^{-2}$.

\section{Application\label{sec:Application}}

The $\beta$ computation presented in Eq.~(\ref{eq:betafinal}) is
not sensitive to $\dot{\bar{\rho}}_{\sigma}$ that is involved in
the definition of the isocurvature perturbation $\delta_{S}$. Instead,
it is a property of quadratic nature of the scalar composite operator
during inflation. Since Eq.~(\ref{eq:betafinal}) does depend on
the masses, in this section, we motivate couple of the mass parameters
from well-motivated nonthermal dark matter models: WIMPZILLAs \cite{Chung:1998zb,Chung:1998ua,Kuzmin:1999zk,Kuzmin:1998kk,Chung:2001cb,Chung:1998bt,Chung:1998is}
and axions \cite{Peccei:1977hh,Weinberg:1977ma,Wilczek:1977pj}. Although
these two particles have different physical origins, they share some
common properties as a cosmological component. Firstly, since they
are massive (at the CMB time at least) and weakly interacting, they
both are good CDM candidates. Also, they can be gravitationally produced
during or after inflation, and this gives rise to isocuvature from
their density perturbations. Furthermore, when their background field
values are negligibly small, the isocurvature perturbation from these
particles is approximated by quadratic form $\sigma^{2}$. In that
case, they would present detectable non-Gaussianties \cite{Chung:2011xd,Kawasaki:2008sn,Hikage:2008sk}
and their cross correlation is characterized by Eq.~(\ref{eq:betafinal}).


\subsection{\label{sub:Weakness}Weakness of $\sigma$ Interactions with $\psi$}

To connect our computation of $\beta$ to observables, a post inflationary
isocurvature scenario is necessary. For the illustrative situations
of axions and WIMPZILLAs, it is sufficient to assume that $\sigma$
has an extremely weak interaction with the reheating degrees of freedom
$\psi$ and the inflaton $\phi$ such that the transfer function of
$\sigma$ is trivial after inflation: with sufficiently small interactions,
$\alpha$ and $\beta$ of Eqs.~(\ref{eq:fractional_powerspectrum})
and (\ref{eq:fractional_cross-correlation}) computed during inflation
can be directly matched without any further transfer function computations
to isocurvature initial condition for CMB codes such as CMBFAST. In
this section, we quantify the requisite weakness of the interactions
and qualitatively discuss the situation when the weakness assumption
is invalid. For example, we will show below that ordinary WIMPs are
too strongly interacting with the reheating degrees of freedom for
this assumption to be valid while axions and WIMPZILLAs are sufficiently
weakly interacting. We also qualitatively describe what extra work
needs to be done to apply this paper for observations in situations
in which the dark matter particles are not extremely weakly interacting.%
\footnote{Because of the cross correlation result in this paper is small, the
discussion here is a bit academic if this discussion applied only
to the cross correlation result. However, the discussion here applies
to the isocurvature 2-point function found in the literature \cite{Seckel:1985tj,Lyth:1991ub,Linde:1996gt,Chung:2004nh,Fox:2004kb,Beltran:2006sq,Hertzberg:2008wr,Chung:2011xd}
which has a realistic chance of being observable in near future experiments.%
}

At the linearized classical equation of motion level, we have the
gauge invariant perturbations $\{\zeta_{j}\}$ being governed by a
linear time evolution operator
\begin{equation}
O[\{\zeta_{j}\}]=0
\end{equation}
where the initial condition for the isocurvature species $j=\sigma$%
\footnote{In our scenario, the isocurvature species stand for the degrees of
freedom constrast with the radiation degrees of freedom.%
} is given by 
\begin{equation}
\zeta_{\sigma}(t_{i})=f(t_{i})
\end{equation}
which in turn is set by the inflationary physics. For example, the
initial time $t_{i}$ can be set to be the time of end of inflation.
The final $\zeta_{\sigma}(t_{f})$ will contain contribution which
does not vanish in the limit $f\rightarrow0$. Hence, one can write
\begin{equation}
\zeta_{\sigma}(t_{f})=G_{t_{f}}^{\sigma}[f(t_{i}),0]+G_{t_{f}}^{\sigma}[0,\zeta_{j\neq\sigma}(t_{i})]
\end{equation}
where $G_{t_{f}}^{\sigma}[D]$ is the $\sigma$ component of the Green's
function derived from the linear operator $O$ which takes the initial
data $D$ and maps it to the final value of $\zeta_{\sigma}(t_{f})$.
Note that we have implicitly assumed the boundary condition such that
$G_{t_{f}}^{\sigma}[0,0]=0$ which means that $G_{t_{f}}^{\sigma}[f(t_{i}),0]$
vanishes as $f(t_{i})\rightarrow0$. 

Now, we will consider two situations in which bound the picture of
super weakly interacting scenarios. In the first scenario, the thermal
plasma generated by the inflaton decay will interact with $\sigma$
sufficiently strongly to make $\delta_{S}$ mix strongly with $\zeta$.
In the second scenario, the inflaton decay to $\sigma$ directly will
realign $\sigma$ fluctuations during radiation domination to those
of $\zeta$, even though $\sigma$ and reheating products are not
interacting appreciably.

First, consider the effects of radiation dominated thermal plasma
on $\sigma$. The mixing rate governing $G_{t_{f}}^{\sigma}[0,\zeta_{j\neq\sigma}(t_{i})]$
is the production rate of $\sigma$ particles from the thermal plasma.
Typically a single channel involving particle $y$ dominates the production
of the $\sigma$ particle from the plasma. (If there are more channels,
the discussion below can easily be generalized.) We thus expect a
qualitative behavior of
\begin{equation}
G_{t_{f}}^{\sigma}[0,\zeta_{j\neq\sigma}(t_{i})]\sim\left(1+\tanh\left[\frac{\Gamma(yy\rightarrow\sigma\sigma,t_{\mbox{max}})}{H(t_{\mbox{max}})}\right]\right)\zeta_{y}
\end{equation}
where $\Gamma(yy\rightarrow\sigma\sigma,t_{\mbox{max}})$ is the reaction
rate for this process at the time that the production rate is maximum
(in $\Gamma(yy\rightarrow\sigma\sigma,t)$ is maximum at $t=t_{\mbox{max}}$
where $t_{\mbox{max}}\in[t_{i},t_{f}]$), $H$ is the expansion rate,
and $\zeta_{y}=O(\zeta_{\mbox{tot}})$.

Hence, one sees that the information about the isocurvature perturbations
depend not only on 
\[
\Gamma(yy\rightarrow\sigma\sigma,t_{\mbox{max}})/H(t_{\mbox{max}})
\]
 but on $t_{f}$ since $t_{\mbox{max }}$ is restricted to be in the
range $t_{\mbox{max}}\in[t_{i},t_{f}].$ For example, the usual CMB
code is run starting with an initial condition at $T\ll T_{\mbox{BBN}}$.
This means that $t_{f}\gg t_{\mbox{BBN}}$ is required to use the
inflationary correlator computations in the CMB code. A general computation
of $G_{t_{f}}^{\sigma}$ needed for the prediction of isocurvature
perturbation effect on CMB temperature is beyond the scope of current
work. To be able to trust the trivial transfer function of
\begin{equation}
G_{t_{f}}^{\sigma}[f(t_{i}),0]\approx f(t_{i})\gg G_{t_{f}}^{\sigma}[0,\zeta_{j\neq\sigma}(t_{i})],
\end{equation}
for superhorizon modes (where $t_{i}$ is say at the end of inflation%
\footnote{\label{fn:moremodecorr}Note that as discussed in footnote \ref{fn:After-inflation-ends},
$\alpha$ can also receive corrections from the departures from the
ideal dS mode function evolution as well as from the time when $m/H$
becomes larger than unity. As discussed there, the quantity $\beta$
is not as sensitive to these corrections. %
}), we can require
\begin{equation}
\frac{\Gamma(yy\rightarrow\sigma\sigma,t_{\mbox{max}})}{H(t_{\mbox{max}})}\ll\frac{\zeta_{\sigma}(t_{i})}{\zeta_{\mbox{tot}}}
\end{equation}
where $t_{\mbox{max}}$ can be at any time between inflation and the
time at which boundary conditions are set for the CMB code. This sets
a bound on the cross section $\langle\sigma v\rangle$ for $yy\rightarrow\sigma\sigma$
to be
\begin{equation}
\langle\sigma v\rangle\ll\frac{\zeta_{\sigma}(t_{i})}{\zeta_{\mbox{tot}}(t_{i})}\frac{g_{*}^{3/4}}{g_{y}}\left(\frac{T_{\mbox{RH}}}{10^{6}\mbox{ GeV}}\right)^{-1}4.2\times10^{-25}\mbox{ GeV}^{-2}\label{eq:cross-section-bound}
\end{equation}
where the bound becomes more stringent for higher reheating temperatures.

This number should be compared to typical thermal WIMP DM candidate
annihilation cross section of $10^{-9}$ GeV$^{-2}$ and a high energy
s-channel scattering at $T_{RH}$ mediated through a vector boson
with a dimensionless coupling $g=\sqrt{4\pi\alpha_{g}}$:
\begin{eqnarray}
\langle\sigma v_{yy\rightarrow A^{\mu}\rightarrow\mbox{light states}}\rangle & \sim & \frac{\alpha_{g}^{2}}{T_{RH}^{2}}\\
 & = & \left(\frac{\alpha}{10^{-1}}\right)^{2}\left(\frac{T_{RH}}{10^{6}\mbox{ GeV}}\right)^{-2}10^{-14}\mbox{ GeV}^{-2}.
\end{eqnarray}
Hence, one sees that WIMP dark matter cannot play the role of the
isocurvature perturbations. That is why if we are to identify our
computation of $\alpha$ and $\beta$ directly to physical observables,
we have to choose the isocurvature degree of freedom to be nonthermal.%
\footnote{Similar arguments can also be made from unitarity \cite{Griest:1989wd}. %
}

Even though the current work applies most immediately without changes
to nonthermal dark matter scenarios having extremely weak interactions,
Eq.~(\ref{eq:cross-section-bound}) is still much bigger than gravity
mediated s-channel interactions
\begin{eqnarray}
\langle\sigma v_{yy\rightarrow g_{\mu\nu}\rightarrow\sigma\sigma}\rangle & \sim & \frac{1}{16\pi^{2}}\frac{T_{RH}^{2}}{M_{p}^{4}}\\
 & \sim & \left(\frac{T_{RH}}{10^{6}\mbox{ GeV}}\right)^{2}10^{-64}\mbox{ GeV}^{-2}.
\end{eqnarray}
For example, axion cross sections for gluon coannihilation behave
as \cite{Kuster:2008zz} 
\begin{eqnarray}
\langle\sigma v_{ag\rightarrow X}\rangle & \sim & \frac{\alpha_{s}^{2}}{8\pi^{2}}\frac{1}{f_{a}^{2}}\label{eq:dummy22}\\
 & \sim & \left(\frac{f_{a}}{10^{12}\mbox{ GeV}}\right)^{-2}10^{-28}\mbox{ GeV}^{-2}\label{eq:dummy22.5}
\end{eqnarray}
where $f_{a}$ is the PQ breaking VEV. Hence, there is a large class
of weakly interacting models for which this work directly applies
without modification. For models for which Eq.~(\ref{eq:cross-section-bound})
is not satisfied, one needs to compute the transfer function associated
with the mixing. Nonetheless, this work will still be useful in setting
the initial conditions for such computations.

Let's see qualitatively what happens when Eq.~(\ref{eq:cross-section-bound})
is not satisfied. In that case, we expect mixing between isocurvature
and curvature perturbations 
\begin{equation}
\zeta_{\sigma}(t_{f})=G_{t_{f}}^{\sigma}[f(t_{i}),0]+G_{t_{f}}^{\sigma}[0,\zeta_{j\neq\sigma}(t_{i})]\sim O(\zeta_{\sigma})+O(\zeta_{\mbox{tot}}).
\end{equation}
Since the curvature perturbations will analogously be 
\begin{equation}
\zeta(t_{f})=G_{t_{f}}^{R}[f(t_{i}),0]+G_{t_{f}}^{R}[0,\zeta_{j\neq\sigma}(t_{i})],
\end{equation}
we would then have
\begin{equation}
\delta_{S}=3\left(\left\{ G_{t_{f}}^{\sigma}[f(t_{i}),0]-G_{t_{f}}^{R}[f(t_{i}),0]\right\} +\left\{ G_{t_{f}}^{\sigma}[0,\zeta_{j\neq\sigma}(t_{i})]-G_{t_{f}}^{R}[0,\zeta_{j\neq\sigma}(t_{i})]\right\} \right).
\end{equation}
Up to the accuracy that all species are equipartitioned, this quantity
may vanish since there is cancellation in each of the terms in the
bracket. It is beyond the scope of the current work to compute more
precisely this cancellation we are focusing on scenarios which satisfy
Eq.~(\ref{eq:cross-section-bound}).

Suppose there is a direct decay of the inflaton to $\sigma$, and
suppose there is no other appreciable interaction between $\sigma$
and other decay products of the inflaton. In that case, it is better
to set the initial time $t_{i}$ to be at the time of inflaton decay
completion such that $G_{t_{f}}^{\sigma}[f(t_{i}),0]$ is still trivial.
In that case, we have 
\begin{eqnarray}
\zeta_{\sigma} & \equiv & -\frac{A}{2}+\frac{\delta\rho_{\sigma}^{\mbox{(grav)}}+\delta\rho_{\sigma}^{(\mbox{decay})}}{3(\bar{\rho}_{\sigma}^{(\mbox{grav})}+\bar{\rho}_{\sigma}^{(\mbox{decay})}+\bar{P}_{\sigma}^{(\mbox{grav})}+\bar{P}_{\sigma}^{(\mbox{decay})})}\\
 & = & r_{\sigma}^{(\mbox{grav})}\zeta_{\sigma}^{(\mbox{grav})}+r_{\sigma}^{(\mbox{decay})}\zeta_{\sigma}^{(\mbox{decay})}
\end{eqnarray}
where $r_{i}$ has been defined in Eq.~(\ref{eq:ratiodef}). Hence,
we have
\begin{eqnarray}
\delta_{S} & = & 3(\zeta_{\sigma}-\zeta_{R})\\
 & = & 3(r_{\sigma}^{(\mbox{grav})}\zeta_{\sigma}^{(\mbox{grav})}+r_{\sigma}^{(\mbox{decay})}\zeta_{\sigma}^{(\mbox{decay})}-\zeta_{R}).
\end{eqnarray}
If $\zeta_{\sigma}^{(\mbox{decay})}=\zeta_{R}$ is assumed, then
\begin{equation}
\delta_{S}=3\left[1-r_{\sigma}^{(\mbox{decay})}\right]\left(\zeta_{\sigma}^{(\mbox{grav})}-\zeta_{R}\right).
\end{equation}
This equation says that if most of the inflaton energy density goes
to $\sigma$, then the isocurvature is negligible.

In the next two subsections, we now consider couple of mass motivations
for nonthermal dark matter isocurvature candidates.

\subsection{WIMPZILLA}

The WIMPZILLA was originally proposed to avoid the restriction from
the assumption that the dark matter is a thermal relic. Thus, the
WIMPZILLA is supposed to either be very heavy and/or very weakly interacting.
In particular, we consider the possibility that the WIMPZILLA is gravitationally
produced during the phase transition out of the quasi-de-Sitter phase
of inflation. In that case, the model is controlled by two parameters:
the ratio of mass to the Hubble scale of inflation $m_{X}/H_{\inf}$,
and the reheating temperature $T_{RH}$, where $X$ denotes a massive
scalar field. Since the energy density is approximated as $\rho_{X}\sim m_{X}^{2}X^{2}$
the relic density of $X$ is estimated as
\begin{equation}
\Omega_{X}h^{2}\sim10^{-1}\left(\frac{H_{e}}{10^{12}GeV}\right)^{2}\left(\frac{T_{RH}}{10^{6}GeV}\right),
\end{equation}
where we have assumed that $m_{X}\sim H_{e}$, because a priori we
know that we can find proper isocurvature and relic density in this
mass range. (For a more detailed discussion of the relic abundance,
see for example \cite{Chung:2011xd}.) The isocurvature power spectrum
depends on the details of the evolution of the background during inflation
because the mode function of massive particle decays as $a^{-3+2\nu}$
(see a related discussion in footnote \ref{fn:moremodecorr}). However,
we can generally obtain $\alpha\sim0.067$ if $m_{X}\lesssim H_{inf}$,
where $H_{inf}$ is the Hubble expansion rate when the CMB scale crosses
the horizon \cite{Chung:2011xd}. The WIMPZILLA isocurvature has also
the quadratic form like the axion. It thus generates the observable
non-Gaussianities estimated as Eq. (\ref{eq:fnl}). Eq.~(\ref{eq:betafinal})
translates to the fractional cross-correlation of
\begin{equation}
\beta_{WIMPZILLA}\approx-0.4\frac{m_{X}}{H_{inf}}\sqrt{\Delta_{\zeta}^{2}}\label{eq:beta_wimpzilla}
\end{equation}
which justifies the constraint used in \cite{Chung:2011xd}. Since
the naive estimate of Eq.~(\ref{eq:betaenhanced}) gives a gross
overestimate $\beta$, one of the merits of this paper is to put such
worries to rest through the proper computation.

\subsection{Axion}

In this subsection we assess the relevance of Eq.~(\ref{eq:betafinal})
to the axion scenario. Firstly, we review the axion scenario. In 1997,
Peccei and Quinn proposed the global $U(1)_{PQ}$ symmetry in order
to solve to the strong CP problem in the QCD. The axion is the Nambu-Goldstone
boson associated with the symmetry after it is broken spontaneously.
Many mechanisms have been proposed to produce axions in the early
universe. We focus only on the ``vacuum misalignment'' mechanism
here following Refs. \cite{Fox:2004kb,Sikivie:2006ni,Preskill:1982cy,Abbott:1982af,Dine:1982ah,Steinhardt:1983ia,Turner:1985si,Hertzberg:2008wr,Kolb:1990vq}.
In early universe, the axions are effectively massless and gain their
mass when the QCD anomaly term (which explicitly breaks PQ symmetry)
becomes physical after the chiral symmetry breaking QCD phase transition.
After the universe cools down and the Hubble friction drops below
the axion mass, the axions begin to coherently oscillate and they
contributes to the CDM component of the universe because of their
long lifetime.

Let us denote the PQ symmetry breaking scale by $f_{a}.$ Because
$n_{a}\propto\theta^{2}$ where $\theta$ is the axion angle, the
relic axion density is estimated as
\begin{equation}
\Omega_{a}h^{2}\sim\begin{cases}
2\times10^{4}\left(\frac{f_{a}/N}{10^{16}\mbox{ GeV}}\right)^{7/6}\left\langle \theta^{2}\right\rangle  & \mbox{for }T_{osc}\gtrsim\Lambda_{QCD}\\
5\times10^{3}\left(\frac{f_{a}/N}{10^{16}\mbox{ GeV}}\right)^{3/2}\left\langle \theta^{2}\right\rangle  & \mbox{for }T_{osc}\lesssim\Lambda_{QCD},
\end{cases}\label{eq:relic_abundance_axion}
\end{equation}
where we have neglected $O(1)$ factors due to diffusion, anharmonic
correction, and temperature-dependent mass correction, and $T_{osc}$
is the temperature at which the axion starts to oscillate. The axion
isocurvature in comoving gauge is written as 
\begin{equation}
\delta_{s}^{(C)}=\omega_{a}\frac{\theta^{2}-\left\langle \theta^{2}\right\rangle }{\left\langle \theta^{2}\right\rangle }=\omega_{a}\frac{2\theta_{i}\delta\theta+\delta\theta^{2}-\left\langle \delta\theta^{2}\right\rangle }{\left\langle \theta^{2}\right\rangle },
\end{equation}
where $\omega_{a}\equiv\Omega_{a}/\Omega_{CDM}$, $\theta_{i}$ is
the average of initial QCD vacuum angle $\theta$ over the observable
universe, and $\delta\theta$ is inhomogeneity of $\theta$, i.e.
$\theta(t,\vec{x})=\theta_{i}(t)+\delta\theta(t,\vec{x})$. Then the
isocurvature power spectrum becomes
\begin{equation}
\widetilde{\left\langle \delta_{s}\delta_{s}\right\rangle }\sim\omega_{a}^{2}\begin{cases}
3.5\times10^{10}\left(\frac{f_{a}/N}{10^{16}\mbox{ GeV}}\right)^{7/3}\widetilde{F} & \mbox{for }f_{a}/N\gtrsim6\times10^{17}GeV\\
2\times10^{9}\left(\frac{f_{a}/N}{10^{16}\mbox{ GeV}}\right)^{3}\widetilde{F} & \mbox{for }f_{a}/N\lesssim6\times10^{17}GeV,
\end{cases}\label{eq:iso_power_axion}
\end{equation}
where
\begin{equation}
\widetilde{F}=4\theta_{i}^{2}\widetilde{\left\langle \delta\theta\delta\theta\right\rangle }+\widetilde{\left\langle \delta\theta^{2}\delta\theta^{2}\right\rangle }+\theta_{i}\left[\langle\delta\theta\delta\theta^{2}\rangle+\langle\delta\theta^{2}\delta\theta\rangle\right].\label{eq:dummy26}
\end{equation}
Since our primary interest is in the cross correlation with $\theta_{i}\approx0$,
we set it to zero.

Therefore, the adiabaticity parameter $\alpha$ defined in Eq. (\ref{eq:fractional_powerspectrum})
is estimated as
\begin{eqnarray}
\alpha & \sim & \omega_{a}^{2}\begin{cases}
1.3\times10^{19}\left(\frac{f_{a}/N}{10^{16}GeV}\right)^{7/3}\Delta_{\theta}^{2} & \mbox{for }f_{a}/N\gtrsim6\times10^{17}GeV\\
8.1\times10^{17}\left(\frac{f_{a}/N}{10^{16}GeV}\right)^{3}\Delta_{\theta}^{2} & \mbox{for }f_{a}/N\lesssim6\times10^{17}GeV,
\end{cases}\label{eq:alpha_axion}\\
\Delta_{\theta}^{2}(p) & = & \frac{p^{3}}{2\pi^{2}}\widetilde{\left\langle \delta\theta^{2}\delta\theta^{2}\right\rangle }\nonumber \\
 & = & \left(\frac{f_{a}}{N}\right)^{-4}\left(\frac{H_{p}^{2}}{2\pi^{2}}\right)^{2}\ln\frac{p}{\Lambda_{IR}},
\end{eqnarray}
where $H_{p}$ is the Hubble scale at the horizon exit of mode $p$,
and $\Lambda_{IR}$ is an IR cut-off. Here we have used Eq.~(\ref{eq:sigma2sigma2})
with the assumption that the axion is effectively massless during
inflation. In the case that $\theta_{i}\ll\delta\theta$, the isocurvature
has the quadratic form of gaussian variable $\delta\theta,$ and it
naturally becomes non-Gaussian perturbation. The isocurvature non-Gaussianity
is estimated as Eq.~(\ref{eq:fnl}).

These parameter constraints and predictions (\ref{eq:relic_abundance_axion}),
(\ref{eq:alpha_axion}) and (\ref{eq:fnl}) already have been investigated
in the literature \cite{Fox:2004kb,Hertzberg:2008wr,Hikage:2008sk,Kawasaki:2008sn}
with the assumption that the axion isocurvature and the curvature
is uncorrelated. Our result from Eq.~(\ref{eq:betafinal}) is 
\begin{equation}
\beta_{axion}=-\frac{\sqrt{\Delta_{\zeta}^{2}}}{2}\left(\ln\frac{p}{\Lambda_{IR}}\right)^{-1/2}\lesssim2.5\times10^{-5}\label{eq:beta_axion}
\end{equation}
which is consistent with the assumptions made in the literature.

\section{Summary\label{sec:Conclusions}}

In this paper, we have presented the first explicit computation of
the gravitational interaction contribution to the cross-correlation
between the curvature and quadratic isocurvature perturbations (which
include dark matter isocurvature candidates such as axions and WIMPZILLAs).
Since the necessary and sufficient condition for the cross-correlation
to dominate over the isocurvature perturbations in the temperature
two-point function is $\left|\beta\right|\gtrsim4\times10^{-2}$,
we have explicitly computed $\beta$, which incidentally is not sensitive
to the background number density of the isocurvature degrees of freedom
and post-inflationary mode function changes on superhorizon scales.
Although a naive estimate of $\beta$ based on a diffeomorphism violating
UV cutoff leads to the possibility of $\beta\sim O(1)$ due to a large
ratio that can appear between the numerator and the denominator of
the expression for $\beta$, our explicitly diffeomorphism invariant
computation leads to $\left|\beta\right|\lesssim\Delta_{\zeta}/2\approx2.5\times10^{-5}$
because the numerator has a suppression as a consequence of a diffeomorphism
Ward identity. Unfortunately, this is far below the current observational
sensitivity of $\left|\beta\right|\gtrsim10^{-2}$. 

The smallness of the cross-correlation is explained by the fact that
the super-horizon mode of the curvature perturbation $\zeta$ can
be smoothly connected to the gauge mode, which is the spatial dilatation,
in the zero external momentum limit. Hence, Eq.~(\ref{eq:sigma2zeta_CG})
vanishes when $p=0$ and $m\neq0$. In other words, this can be seen
as a suppression due to a diffeomorphism Ward identity (i.e. uniform
spatial rescaling invariance). A nontrivial structure revealed through
our explicit computation is the suppression's non-analytic structure
with respect to $p$: the cross correlation cannot be Taylor-expanded
at $p=0$, and this contribution is not $p^{2}/a^{2}$-suppressed. 

Our rigorous result which incorporates UV renormalization of the composite
operator in the curved background is also shown to be consistent with
an estimate based on a soft-$\zeta$ theorem, which allows one to
factorize $\left\langle \zeta\zeta\right\rangle $ from $\left\langle \sigma^{2}\zeta\right\rangle $
as explained in Eq.~(\ref{eq:soft-z_arg-2pt_fn}). However, Eq.~(\ref{eq:soft-z_arg-2pt_fn})
requires two assumptions that can only be justified by an honest computation
such as what is presented in subsection \ref{sub:Two-point-Correlators}:
\begin{enumerate}
\item There is an effective IR cutoff of $p$ in evaluating $\langle\sigma^{2}\rangle$
due to the external momentum $p$ inserted into the composite operator.
\item The only UV renormalization property of $\langle\sigma^{2}\rangle$
that is relevant to leading $\hbar$ approximation is the preservation
of diffeomorphism invariance.
\end{enumerate}
Note that the proper diffeomorphism invariant UV treatment also allowed
us to demonstrate that the cross-correlation is indeed gauge-invariant
with one-loop correction through the gravitational coupling. This
gauge invariance is checked explicitly by computing our cross correlation
in both the comoving gauge and the uniform curvature gauge. 

Physically, the curvature perturbation $\zeta$ can affect the particle
density $\rho_{\sigma}$ and generate correlations only at its horizon
crossing, because $\zeta$ freezes out after its horizon exit, after
which it can be effectively treated as a gauge mode. Positive cross
correlation corresponds to the situation in which the $1+\zeta$ enhancement
in the expansion enhances the particle production (assuming that this
enhances inhomogeneity) while the negative cross correlation corresponds
to the situation in which the $1+\zeta$ enhancement in the expansion
dilutes the particle inhomogeneity. The latter dilution effect leads
to $\beta>0$, while the particle production enhancement effect corresponds
to the quadratic scenario that we were interested in this paper. This
explains the sign $\beta<0$ of our result.

Given the robustness of the smallness of $\beta$, the gravitational
interaction contribution to the cross correlation should be negligible
in most nonthermal dark matter isocurvature scenarios. In addition
to giving a concrete computation that supports this statement, our
work serves as an interesting lesson in computing correlators of composite
operators in curved spacetime in the context of inflationary cosmology.

\section{Acknowledgments}

This work was supported in part by the DOE through grant DE-FG02-95ER40896
and WARF. We thank the hospitality and support of KIAS where part
of this work was accomplished. DJHC thanks Lisa Everett for comments
on the manuscript.

\appendix

\section{Behaviors of Transfer functions for Adiabatic and Isocurvature initial
condition\label{sec:Behaviors-of-Transfer}}

The CMB temperature fluctuation with the leading order approximation
(the integrated Sachs-Wolfe term is neglected) in the Newtonian gauge
($B=F=0$, $E=2\Phi$, $A=-2\Psi$) is
\begin{equation}
\frac{\Delta T}{T}\approx\frac{1}{4}\left.\delta_{\gamma}\right|_{r}+\left.\Phi\right|_{r},\label{eq:SW_term}
\end{equation}
where the perturbations on the rhs are evaluated at the recombination.
We can obtain these perturbations by solving the Einstein and Boltzmann
equations with given initial conditions. A projection from a given
initial condition to the final CMB temperature fluctuation is called
transfer function. In the following subsections, we calculate that
the $k$-dependence of the transfer functions for the adiabatic and
the isocurvature initial conditions. In particular, we show that the
isocurvature transfer function has the additional suppression factor
$k_{eq}/k$ compared to the adiabatic one for small scale $k\gg k_{eq}$.
Here we basically follow the calculation by Ref. \cite{Dodelson:2003ft,Mukhanov:2003xr}.

\subsection{Perturbation Equations}

For explicit computation, we choose the Newtonian gauge for the scalar
metric perturbation (\ref{eq:deltag_s}). For simplicity, we consider
only photon and CDM fluids, which are denoted in the following equations
by subscript $\gamma$ and $m$, respectively. This assumption is
valid for the sake of identifying the difference between transfer
functions for adiabatic and isocurvature initial conditions, although
baryon and neutrino should be taken into account for accurate description
for transfer functions.

The conservation equations for dark matter and photon fluids in Fourier
space are
\begin{eqnarray}
\delta_{m}' & = & k^{2}V_{m}+3\Psi,\label{eq:Ddm}\\
V_{m}' & = & -\mathcal{H}V_{m}-\Phi,\\
\delta_{\gamma}' & = & \frac{4}{3}k^{2}V_{\gamma}+4\Psi',\label{eq:Ddr}\\
V_{\gamma}' & = & -\frac{1}{4}\delta_{\gamma}-\Phi,\label{eq:DVr}
\end{eqnarray}
where ' denotes the time derivative with respect to conformal time
$\eta$, $\mathcal{H}\equiv a'/a$, $\delta_{a}\equiv\delta\rho_{a}/\rho_{a}$.
Note that $\Phi=\Psi$ since they are perfect fluids. $V_{X}$ is
the peculiar velocity for fluid $X$. These four equation are combined
by eliminating $V_{X}$, and we have
\begin{eqnarray}
\left(a\left(\delta_{m}'-3\Phi'\right)\right)' & = & ak^{2}\Phi,\label{eq:Evol_dm}\\
\delta_{\gamma}'' & = & 4\Phi''-\frac{k^{2}}{3}\left(\delta_{\gamma}+4\Phi\right).\label{eq:Evol_dr}
\end{eqnarray}
The evolution of the metric perturbation is encoded in the Einstein
equations. (00) and (ii) components are
\begin{eqnarray}
k^{2}\Phi+3\mathcal{H}\left(\Phi'+\mathcal{H}\Phi\right) & = & -\frac{1}{2M_{p}^{2}}a^{2}\left(\rho_{m}\delta_{m}+\rho_{\gamma}\delta_{\gamma}\right),\label{eq:Einstein00}\\
\Phi''+3\mathcal{H}\Phi'+\left(2\frac{a''}{a}-\mathcal{H}^{2}\right)\Phi & = & \frac{1}{6M_{p}^{2}}a^{2}\rho_{\gamma}\delta_{\gamma.}\label{eq:Einsteinii}
\end{eqnarray}
Combining with other components, we also find the Poisson equation
\begin{eqnarray}
-k^{2}\Phi & = & \frac{3}{2}\mathcal{H}^{2}\left[\Omega_{m}\delta_{m}+\Omega_{\gamma}\delta_{\gamma}-3\mathcal{H}\left(\Omega_{m}V_{m}+\frac{4}{3}\Omega_{\gamma}V_{\gamma}\right)\right].\label{eq:Poisson_eq}
\end{eqnarray}

With the definition of isocurvature (\ref{eq:deltaS}) in Section
\ref{sec:Curvature-and-Isocurvature}
\begin{equation}
\delta_{S}=\delta_{m}-\frac{3}{4}\delta_{\gamma},\label{eq:Def_Isoc}
\end{equation}
 where we have used $p_{\gamma}=\rho_{\gamma}/3$ and $p_{m}=0$,
we rewrite the differential equations of fluid and metric perturbations
in terms of $\Phi$ and $\delta_{S}$
\begin{eqnarray}
\Phi''+3\mathcal{H}\left(1+c_{s}^{2}\right)\Phi'+\left[2\mathcal{H}'+\mathcal{H}^{2}\left(1+3c_{s}^{2}\right)\right]\Phi+k^{2}c_{s}^{2}\Phi & = & -\frac{2}{3}\frac{c_{s}^{2}}{M_{p}^{2}}a^{2}\rho_{m}\delta_{S},\label{eq:Evol_Phi}\\
\frac{1}{3c_{s}^{2}}\delta_{S}''+\frac{a'}{a}\delta_{S}'+\frac{k^{2}y}{4}\delta_{S} & = & -\frac{1}{6}y^{2}k^{4}\tau_{eq}^{2}\Phi,\label{eq:Evol_dS}
\end{eqnarray}
where 
\begin{equation}
y\equiv\frac{a}{a_{eq}}=\frac{\rho_{m}}{\rho_{\gamma}},\quad\tau_{eq}=\frac{\sqrt{2}}{a_{eq}H_{eq}},\quad c_{s}^{-2}\equiv3\left(1+\frac{3}{4}y\right).
\end{equation}
In $\eta\to0$ limit, Eqs. (\ref{eq:Evol_Phi}) and (\ref{eq:Evol_dS})
admit two linearly independent solutions $\Phi(k,\eta\to0)=\Phi^{i}(k),\,\delta_{S}(k,\eta\to0)=0$,
and $\Phi(k,\eta\to0)=0,\,\delta_{S}(k,\eta\to0)=\delta_{S}^{i}(k)$,
which corresponds to adiabatic initial condition and isocurvature
initial condition, respectively.

\subsection{Adiabatic Initial Condition}

For large scale perturbations, which enters the horizon later than
the recombination. $\delta_{S}$ remains zero according to Eq. (\ref{eq:Evol_dS}),
and thus Eq. (\ref{eq:Evol_Phi}) is rewritten as
\begin{equation}
\frac{d^{2}\Phi}{dy^{2}}+\frac{21y^{2}+54y+32}{2y(y+1)(3y+4)}\frac{d\Phi}{dy}+\frac{\Phi}{y(y+1)(3y+4)}=0,
\end{equation}
where is called as Kodama-Sasaki equation. This differential equation
can be exactly solved, and we find 
\begin{equation}
\Phi(k_{l},y\gg1)=\frac{9}{10}\Phi^{i}(k_{l}),
\end{equation}
where the subscript $l$ stands for ``super-horizon''. For photon
energy density $\delta_{\gamma}$, Eq. (\ref{eq:Ddr}) in the long
wavelength limit yields
\begin{equation}
\frac{1}{4}\delta_{\gamma}-\Phi=const.
\end{equation}
and also Eq. (\ref{eq:Einstein00}) gives 
\begin{equation}
\delta_{\gamma}(k_{l,}\eta\to0)=-2\Phi(k_{l},\eta)=-2\Phi^{i}(k_{l}).
\end{equation}

For small scale perturbation, which enter the horizon during the radiation
dominated(RD) era, in the early RD limit $\eta\ll\eta_{eq}$, Eq.
(\ref{eq:Evol_Phi}) becomes
\begin{equation}
\Phi''+\frac{4}{\eta}\Phi'+\frac{k^{2}}{3}\Phi=0,\label{eq:Phi_sub_init}
\end{equation}
and its solution with the adiabatic initial condition
\begin{equation}
\Phi(k_{s},\eta<\eta_{eq})=\frac{3}{\left(w\eta\right)^{3}}\left(\sin w\eta-w\eta\cos w\eta\right)\Phi^{i}(k_{s}),
\end{equation}
where $w=k/\sqrt{3}$. After the perturbation enters the horizon,
\begin{eqnarray}
\Phi(k_{s},\eta<\eta_{eq}) & \approx & -\frac{3\cos w\eta}{\left(w\eta\right)^{2}}\Phi^{i}(k_{s}),\\
\delta_{\gamma}(k_{s},\eta<\eta_{eq}) & \approx & -\frac{2M_{p}^{2}}{\rho_{\gamma}a^{2}}\Phi(k_{s},\eta)=6\Phi^{i}(k_{s})\cos w\eta,\label{eq:dr_in_sub_RD}
\end{eqnarray}
where the subscript $s$ means ``sub-horizon'', and the second equation
is obtained by the Poisson equation (\ref{eq:Poisson_eq}). Plugging
this solution into Eq. (\ref{eq:Evol_dm}), we find that 
\begin{equation}
\delta_{m}(k_{s},\eta<\eta_{eq})\approx-9\Phi^{i}(k_{s})\left(\ln w\eta+\gamma-\frac{1}{2}\right),\label{eq:dm_in_RD}
\end{equation}
where $\gamma$ is the Euler Gamma constant. This shows that the dark
matter density perturbation grows logarithmically during the RD era.

Now we should match this with the solutions in the matter dominated(MD)
era. Because the time derivatives of $\Phi$ is negligible compared
to the spatial derivatives, Eq. (\ref{eq:Evol_dm}) is approximated
as 
\begin{equation}
\delta_{m}''+\mathcal{H}\delta_{m}'\approx-k^{2}\Phi\approx\frac{3}{2}\mathcal{H}^{2}\Omega_{m}\delta_{m},
\end{equation}
where we have used the Poisson equation (\ref{eq:Poisson_eq}). Then,
it is rewritten as
\begin{equation}
y(1+y)\frac{d^{2}\delta_{m}}{d^{2}y}+\left(1+\frac{3}{2}y\right)\frac{d\delta_{m}}{dy}-\frac{3}{2}\delta_{m}=0,
\end{equation}
and its general solution is 
\begin{equation}
\delta_{m}=c_{1}\left(1+\frac{3}{2}y\right)+c_{2}\left[\left(1+\frac{3}{2}y\right)\ln\frac{\sqrt{1+y}+1}{\sqrt{1+y}-1}-3\sqrt{1+y}\right].\label{eq:dm_gen_sol_MD}
\end{equation}
Matching this solution with Eq. (\ref{eq:dm_in_RD}) at $y\ll1$,
we find
\begin{eqnarray}
\delta_{m}(k_{s},\eta>\eta_{eq}) & = & -9\Phi^{i}(k)\left(\ln2w\eta_{*}+\gamma-\frac{7}{2}\right)\left(1+\frac{3}{2}y\right)+9\Phi^{i}(k)\left[\left(1+\frac{3}{2}y\right)\ln\frac{\sqrt{1+y}+1}{\sqrt{1+y}-1}-3\sqrt{1+y}\right],\nonumber \\
 & \to & -\frac{27}{2}y\Phi^{i}(k)\left(\ln2w\eta_{*}+\gamma-\frac{7}{2}\right)\;\mbox{when }y\gg1.\label{eq:dm_sub_MD}
\end{eqnarray}
where $\eta_{*}\equiv\eta_{eq}/\left(\sqrt{2}-1\right)=2\tau_{eq}.$
Note that we have used the results from the Friedman equation 
\begin{eqnarray}
\mathcal{H}^{2} & = & \frac{a_{eq}^{2}H_{eq}^{2}}{2}\left(\frac{1}{y}+\frac{1}{y^{2}}\right),\\
y & = & \frac{\eta^{2}}{\left(2\tau_{eq}\right)^{2}}+\frac{\eta}{\tau_{eq}},
\end{eqnarray}
and Eq. (\ref{eq:dm_sub_MD}) corresponds to Eq. (150) in Ref. \cite{Mukhanov:2003xr}.

Then using Eqs. (\ref{eq:Poisson_eq}) and (\ref{eq:dm_sub_MD}),
we get 
\begin{equation}
\Phi(k_{s},\eta>\eta_{eq})\approx\frac{\ln\left(0.15k_{s}\eta_{eq}\right)}{\left(0.27k_{s}\eta_{eq}\right)^{2}}\Phi^{i}(k_{s}).
\end{equation}
 This shows that the gravitational potential is frozen after the matter-radiation
equality. Similarly, we first find the general solution of Eq. (\ref{eq:Evol_dr})
for sub-horizon modes
\begin{equation}
\delta_{\gamma}=c_{1}\cos w\eta+c_{2}\sin w\eta-4\Phi,\label{eq:dr_gen_sol_MD}
\end{equation}
where we have neglected that time derivatives of $\Phi$. Then matching
this with Eq. (\ref{eq:dr_in_sub_RD}), we get 

\begin{equation}
\delta_{\gamma}(k_{s},\eta>\eta_{eq})\approx\left[6\cos\left(w\eta\right)-4\frac{\ln\left(0.15k_{s}\eta_{eq}\right)}{\left(0.27k_{s}\eta_{eq}\right)^{2}}\right]\Phi^{i}(k_{s}).
\end{equation}
Now we return factors due to the Silk damping and the acoustic sound
speed

\begin{equation}
\delta_{\gamma}(k_{s},\eta>\eta_{eq})\approx\left[3^{5/4}\sqrt{4c_{s}}\cos\left(k_{s}\int^{\eta}c_{s}(\eta')d\eta'\right)e^{-(k_{s}/k_{D})^{2}}-\frac{4}{3c_{s}^{2}}\frac{\ln\left(0.15k_{s}\eta_{eq}\right)}{\left(0.27k_{s}\eta_{eq}\right)^{2}}\right]\Phi^{i}(k_{s}),
\end{equation}
which is Eq. (153) in Ref. \cite{Mukhanov:2003xr}. Notice that the
the first term is dominant for the scales we are interested in. However,
the second term becomes important for very small scales where the
diffusion damping is not negligible, $k\gtrsim k_{D}.$ 

Finally, the SW term (\ref{eq:SW_term}) becomes
\begin{eqnarray}
\frac{\Delta T}{T} & \approx & \begin{cases}
6\Phi^{i}(k)\cos w\eta & \mbox{if }k>k_{eq}\\
\frac{3}{10}\Phi^{i}(k) & \mbox{if }k<\eta_{r}^{-1}.
\end{cases}\label{eq:SW_adiabatic}
\end{eqnarray}
Note that 
\begin{equation}
\zeta^{i}\approx\zeta_{R}^{i}=-\Phi^{i}+\frac{1}{4}\delta_{\gamma}^{i}=-\frac{3}{2}\Phi^{i}.
\end{equation}

\subsection{Isocurvature initial condition}

For large scale perturbations, $\delta_{S}$ remains constant, and
Eq. (\ref{eq:Evol_Phi}) has the solution
\begin{equation}
\Phi(k_{l},\eta)=-\left(\frac{x}{5}\right)\frac{x^{2}+6x+10}{\left(x+2\right)^{3}}\delta_{S}^{i}(k_{l}),\label{eq:Phi_in_MD}
\end{equation}
where $x\equiv\eta/\eta_{eq}$. In the MD era, Eq. (\ref{eq:Phi_in_MD})
gives
\begin{equation}
\Phi(k_{l},\eta\gg\eta_{eq})=-\frac{1}{2}\delta_{m}(k_{l},\eta\gg\eta_{eq})=\frac{1}{4}\delta_{\gamma}(k_{l},\eta\gg\eta_{eq})=-\frac{1}{5}\delta_{S}^{i}(k_{l}),
\end{equation}
where the last two equations are obtained from Eq. (\ref{eq:Einstein00}).

Now, we will see how the perturbations evolve during the RD era, and
how they are connected small scale perturbations. In the early RD
era, the source term and the last term on the left hand side of Eq.
(\ref{eq:Evol_dS}) is negligible because they are higher order in
$y$. Thus, the solution $\delta_{S}$ remains constant even inside
the horizon. In that case, Eq. (\ref{eq:Evol_Phi}) becomes Eq. (\ref{eq:Phi_sub_init})
with the source term $\delta_{S}/2y\eta_{eq}^{2}$. Then we find its
solution that matches with Eq. (\ref{eq:Phi_in_MD})
\begin{equation}
\Phi(k,\eta<\eta_{eq})=-\frac{\eta}{\eta_{eq}}\frac{1}{\left(w\eta\right)^{4}}\left[1+\frac{\left(w\eta\right)^{2}}{2}-\left(\cos w\eta+w\eta\sin w\eta\right)\right]\delta_{S}^{i}(k).\label{eq:Phi_in_early_RD}
\end{equation}
Furthermore, in the $w\eta\to0$ limit, we have 
\begin{equation}
\Phi(k_{l},\eta<\eta_{eq})\approx-\frac{1}{8}\delta_{S}^{i}(k_{l})\left(1-\frac{\left(w\eta\right)^{2}}{18}\right)y,
\end{equation}
and putting this into Eq. (\ref{eq:Einstein00}), we find that 
\begin{eqnarray}
\delta_{\gamma}(k_{l},\eta<\eta_{eq}) & \approx & -\frac{1}{2}\delta_{S}^{i}(k_{l})\left(1-\frac{7}{18}\left(w\eta\right)^{2}\right)y,\\
\delta_{m}(k_{l},\eta<\eta_{eq}) & \approx & \delta_{S}^{i}(k_{l})\left(1-\frac{3}{8}y\right)+\frac{7}{48}\delta_{S}^{i}(k)y\left(w\eta\right)^{2}.
\end{eqnarray}
As explained in Section \ref{sub:Observational-Constraints-on-Isocurvature},
we have that $\Phi$ and $\delta_{\gamma}$ grows like $a$ during
the RD era, meanwhile $\delta_{m}$ decreases. 

For sub-horizon modes, Eq. (\ref{eq:Phi_in_early_RD}) becomes 
\begin{eqnarray}
\Phi(k_{s},\eta<\eta_{eq}) & \approx & -\frac{y}{\left(w\eta\right)^{3}}\left(\frac{w\eta}{2}-\sin w\eta\right)\delta_{S}^{i}(k_{s}),
\end{eqnarray}
and again plugging this into Eq. (\ref{eq:Einstein00}) yields
\begin{eqnarray}
\delta_{m}(k_{s},\eta<\eta_{eq}) & \approx & -\left(\frac{3}{2}\frac{\sin w\eta}{w\eta}y-1\right)\delta_{S}^{i}(k_{s}),\\
\delta_{\gamma}(k_{s},\eta<\eta_{eq}) & \approx & -\frac{2\sin w\eta}{w\eta}y\delta_{S}^{i}(k_{s}).
\end{eqnarray}
Matching these with general solutions of perturbations (\ref{eq:dm_gen_sol_MD})
and (\ref{eq:dr_gen_sol_MD}), and also using Poisson equation (\ref{eq:Poisson_eq})
in the MD era, we get
\begin{eqnarray}
\delta_{m}(k_{s},\eta>\eta_{eq}) & \approx & \left(1+\frac{3}{2}y\right)\delta_{S}^{i}(k_{s}),\\
\delta_{\gamma}(k_{s},\eta>\eta_{eq}) & \approx & \left[-\frac{1}{0.35k_{s}\eta_{eq}}\sin\left(w\eta\right)+4\frac{1}{\left(0.8k_{s}\eta_{eq}\right)^{2}}\right]\delta_{S}^{i}(k_{s}),\\
\Phi(k_{s},\eta>\eta_{eq}) & \approx & -\frac{1}{\left(0.8k_{s}\eta_{eq}\right)^{2}}\delta_{S}^{i}(k_{s}),
\end{eqnarray}

Then the SW term becomes
\begin{eqnarray}
\frac{\Delta T}{T} & \approx & \begin{cases}
-\frac{1}{0.35k\eta_{eq}}\delta_{S}^{i}(k)\sin\left(w\eta\right) & \mbox{if }k>k_{eq}\\
-\frac{2}{5}\delta_{S}^{i}(k) & \mbox{if }k<\eta_{r}^{-1}.
\end{cases}\label{eq:SW_isocurvature}
\end{eqnarray}
Now we see from Eqs. (\ref{eq:SW_adiabatic}) and (\ref{eq:SW_isocurvature})
that the isocurvature transfer function has the additional suppression
factor $k_{eq}/k$ compared to the adiabatic one for small scale $k>k_{eq}$.

\section{Review of Diffeomorphism Invariance\label{sec:Review-of-Diffeomorphism}}

A symmetry in a classical field theory is preserved at the quantum
level, if the regulator preserves this symmetry and if the functional
measure is invariant under the symmetry transformation. The quantum
symmetry is reflected in the transformation of the correlation functions.

For example, consider a scalar field $\sigma$ on a fixed manifold
$(\mathcal{M},g)$. The two point function is 
\begin{equation}
\langle\sigma(x)\sigma(y)\rangle_{g}=\int D\phi e^{iS(\sigma;g)}\sigma(x)\sigma(y)
\end{equation}
 The two point function only depends on the metric field $g$ and
points $x,y$. Intuitively, the symmetry says for any diffeomorphism
$\varphi:\mathcal{M}\mapsto\mathcal{M}$, the metric field and the
points changes as 
\begin{equation}
g\mapsto\tilde{g}=(\varphi^{-1})^{*}g,\, x\mapsto\tilde{x}=\varphi(x),\, y\mapsto\tilde{y}=\varphi(y)
\end{equation}
 then the two-point function should remain invariant, i.e. 
\begin{equation}
\langle\sigma(x)\sigma(y)\rangle_{g}=\langle\sigma(\tilde{x})\sigma(\tilde{y})\rangle_{\tilde{g}}.\label{eq:Ward-1}
\end{equation}
 The Ward identity is the infinitesimal version of this relation.

Let $\varphi=\exp(\epsilon X)$, then 
\begin{eqnarray}
\tilde{g} & = & \exp(-\epsilon X)^{*}g=g-\epsilon\mathcal{L}_{X}g+\cdots\\
S(\tilde{g},\sigma) & = & S(g,\sigma)-\epsilon\int d^{4}x\sqrt{g}\frac{1}{2}T_{\sigma}^{\mu\nu}\mathcal{L}_{X}(g)_{\mu\nu}+\cdots\\
\sigma(\tilde{x}) & = & \sigma(x)+\epsilon\mathcal{L}_{X}\sigma(x)+\cdots
\end{eqnarray}
 Plugging this into Eq. (\ref{eq:Ward-1}) and Taylor expand with
respect to $\epsilon$, one get 
\begin{equation}
-i\int d^{4}z\sqrt{g}\frac{1}{2}\mathcal{L}_{X}(g)_{\mu\nu}(z)\langle T_{z}^{\mu\nu}\sigma_{x}\sigma_{y}\rangle_{g}+\langle\mathcal{L}_{X}(\sigma)_{x}\sigma_{y}\rangle_{g}+\langle\sigma_{x}\mathcal{L}_{X}(\sigma)_{y}\rangle_{g}=0.
\end{equation}
 Or equivalently, using 
\begin{equation}
\mathcal{L}_{X}(g)_{\mu\nu}=\nabla_{\mu}X_{\nu}+\nabla_{\nu}X_{\mu}
\end{equation}
 and perform integration by part, we obtain 
\begin{equation}
i\nabla_{\mu}\langle T_{z}^{\mu\nu}\sigma_{x}\sigma_{y}\rangle_{g}=\frac{1}{\sqrt{g_{x}}}\delta^{4}(x-z)g^{\alpha\nu}\frac{\partial}{\partial x^{\alpha}}\langle\sigma_{x}\sigma_{y}\rangle_{g}+\frac{1}{\sqrt{g_{y}}}\delta^{4}(y-z)g^{\alpha\nu}\frac{\partial}{\partial y^{\alpha}}\langle\sigma_{x}\sigma_{y}\rangle_{g}\label{eq:Ward-ID}
\end{equation}
which is the Ward identity for the path ordered vacuum expectation
value. We can then write down the in-in expectation value Ward identity
as 
\begin{eqnarray}
i\nabla_{\mu}\langle in|T_{z}^{\mu\nu+}\sigma_{x}^{+}\sigma_{y}^{+}|in\rangle_{g} & = & \frac{1}{\sqrt{g_{x}}}\delta^{4}(x-z)g_{x}^{\alpha\nu}\frac{\partial}{\partial x^{\alpha}}\langle in|\sigma_{x}^{+}\sigma_{y}^{+}|in\rangle_{g}\nonumber \\
 &  & +\frac{1}{\sqrt{g_{y}}}\delta^{4}(y-z)g_{y}^{\alpha\nu}\frac{\partial}{\partial y^{\alpha}}\langle in|\sigma_{x}^{+}\sigma_{y}^{+}|in\rangle_{g}\label{eq:Ward-ID-inin-1}\\
i\nabla_{\mu}\langle in|T_{z}^{\mu\nu-}\sigma_{x}^{+}\sigma_{y}^{+}|in\rangle_{g} & = & 0\label{eq:Ward-ID-inin-2}
\end{eqnarray}
where we kept the external operator inserted on the forward branch.
The fact that Eq. (\ref{eq:Ward-ID-inin-2}) has no contact term is
easy to understand, since $T_{z}^{\mu\nu-}$ is inserted on the backward
time branch of the manifold, it can never \emph{contact} points $x$
and $y$.

\section{ADM formalism and Interaction Hamiltonian\label{sec:ADM-formalism}}

We consider an inflationary model with the inflaton $\phi$ and an
extra free massive scalar $\sigma$, where $\sigma$ is only gravitationally
coupled with $\phi$. 
\begin{equation}
S=\int(dx)\frac{1}{2}M_{p}^{2}R+[-\frac{1}{2}g^{\mu\nu}\partial_{\mu}\phi\partial_{\nu}\phi-V(\phi)]+[-\frac{1}{2}g^{\mu\nu}\partial_{\mu}\sigma\partial_{\nu}\sigma-U(\sigma)]\label{eq:action_S}
\end{equation}
where $M_{p}^{2}=\frac{1}{8\pi G}=1$ and $(dx)=d^{4}x\sqrt{|\det(g_{\mu\nu})|}$.
The metric can be parametrized using ADM formalism \cite{Arnowitt:1962}%
\footnote{We use $(-+++)$ sign convention for the metric, and physical time
$t$ .%
}, 
\begin{equation}
g_{\mu\nu}=\left(\begin{array}{cc}
-N^{2}+h_{ij}N^{i}N^{j} & h_{ij}N^{j}\\
h_{ij}N^{j} & h_{ij}
\end{array}\right),\quad g^{\mu\nu}=\left(\begin{array}{cc}
-N^{-2} & N^{i}N^{-2}\\
N^{i}N^{-2} & h^{ij}-N^{i}N^{j}N^{-2}
\end{array}\right),
\end{equation}
 where $h_{ij}$ is the metric tensor on the constant time hyper-surface,
and $h^{ij}$ is the inverse metric. We use Latin indices $i,j\cdots$
for objects on the 3-dimensional constant time hyper-surface, and
we use $h_{ij}$ and $h^{ij}$ to raise and lower the indices. Then
the action (\ref{eq:action_S}) is rewritten as 
\begin{eqnarray}
S & = & \frac{1}{2}\int(dx)\sqrt{h}\left[NR^{(3)}-2NV(\phi)-2NU(\sigma)+N^{-1}\left(E_{ij}E^{ij}-E^{2}\right)+N^{-1}\left(\dot{\phi}-N^{i}\partial_{i}\phi\right)^{2}-Nh^{ij}\partial_{i}\phi\partial_{j}\phi\right.\label{eq:action_S_adm}\\
 &  & \left.+N^{-1}\left(\dot{\sigma}-N^{i}\partial_{i}\sigma\right)^{2}-Nh^{ij}\partial_{i}\sigma\partial_{j}\sigma\right],\nonumber 
\end{eqnarray}
 where $E_{ij}$ and $E$ are given by 
\begin{eqnarray}
E_{ij} & = & \frac{1}{2}(\dot{h}_{ij}-\nabla_{i}^{(3)}N_{j}-\nabla_{j}^{(3)}N_{i}).\\
E & = & E_{ij}h^{ij}.
\end{eqnarray}

Consider the background solution driven by the inflaton, 
\begin{equation}
\phi^{(0)}=\bar{\phi}(t),\quad\sigma^{(0)}=0,\quad g_{\mu\nu}^{(0)}=\left(\begin{array}{cc}
-1 & 0\\
0 & a^{2}(t)\delta_{ij}
\end{array}\right),
\end{equation}
where they satisfy the background equations of motion 
\begin{eqnarray}
3H^{2} & = & \frac{1}{2}\dot{\bar{\phi}}^{2}+V(\bar{\phi})\\
\dot{H} & = & -\frac{1}{2}\dot{\bar{\phi}}^{2}\\
\ddot{\bar{\phi}}+3H\dot{\bar{\phi}}+V'(\bar{\phi}) & = & 0.
\end{eqnarray}
The action for the perturbations can be obtained by Taylor-expanding
the full action around the background solution. However, we may reduce
the number of variables by imposing the ADM constraints: 
\begin{eqnarray}
0 & = & \frac{1}{N}[R^{(3)}-\frac{1}{N^{2}}(E_{ij}E^{ij}-E^{2})]-2NT^{00}\\
0 & = & \frac{2}{N}\nabla_{i}^{(3)}[\frac{1}{N}(E^{ij}-Eh^{ij})]+2N^{j}T^{00}+2T^{0j}
\end{eqnarray}
where 
\begin{eqnarray}
T^{\mu\nu} & = & T_{\phi}^{\mu\nu}+T_{\sigma}^{\mu\nu},\\
T_{\phi}^{\mu\nu} & = & -g^{\mu\nu}\left[\frac{1}{2}\left(\partial\phi\right)^{2}+V(\phi)\right]+\partial^{\mu}\phi\partial^{\nu}\phi,\\
T_{\sigma}^{\mu\nu} & = & -g^{\mu\nu}\left[\frac{1}{2}\left(\partial\sigma\right)^{2}+U(\sigma)\right]+\partial^{\mu}\sigma\partial^{\nu}\sigma,
\end{eqnarray}
and choose a gauge. 

One commonly used gauge is the comoving gauge, defined by %
\footnote{In this section, Latin indices $i,j$ are raised and lowered by $\delta_{ij}$,
and repeated indices are contracted.%
} 
\begin{equation}
\delta\phi=0,\quad\gamma_{ii}=0,\quad\partial_{i}\gamma_{ij}=0
\end{equation}
where 
\begin{equation}
h_{ij}=a^{2}(t)[e^{\Gamma}]_{ij},\quad\Gamma_{ij}=2\zeta\delta_{ij}+\gamma_{ij}\label{eq:h_ij}
\end{equation}
The solution of $N$ and $N^{i}$ is
\begin{equation}
N^{(1,C)}=\frac{\dot{\zeta}}{H},\quad N_{i}^{(1,C)}=\partial_{i}[-\frac{\zeta}{H}+\epsilon\frac{a^{2}}{\nabla^{2}}\dot{\zeta}].
\end{equation}
We find the scalar metric perturbations are 
\begin{equation}
\delta g_{\mu\nu}^{(C)}=\left(\begin{array}{cc}
-2\frac{\dot{\zeta}}{H} & (-\frac{\zeta}{H}+\epsilon\frac{a^{2}}{\nabla^{2}}\dot{\zeta})_{,i}\\
(-\frac{\zeta}{H}+\epsilon\frac{a^{2}}{\nabla^{2}}\dot{\zeta})_{,i} & a^{2}\delta_{ij}2\zeta
\end{array}\right),\label{eq:del_g_C}
\end{equation}
where $\epsilon\equiv\dot{H}/H^{2}.$ Plugging in the linear metric
perturbation back to the action (\ref{eq:action_S_adm}), we can get
the perturbed action action up to cubic order 
\begin{eqnarray}
S^{(C)} & = & S_{\zeta\zeta}^{(C)}+S_{\sigma\sigma}^{(C)}+S_{\gamma\gamma}^{(C)}+S_{\zeta\zeta\zeta}^{(C)}+S_{\zeta\sigma\sigma}^{(C)}+\cdots
\end{eqnarray}
where 
\begin{eqnarray}
S_{\zeta\zeta}^{(C)} & = & \int dtd^{3}xa_{x}^{3}\epsilon(\dot{\zeta}^{2}-(\frac{\nabla}{a}\zeta)^{2})\label{eq:free-zeta-action}\\
S_{\zeta\sigma\sigma}^{(C)} & = & \int d^{4}xa_{x}^{3}[T_{\sigma}^{ij}a^{2}\delta_{ij}\zeta+T_{\sigma}^{0i}(-\frac{\zeta}{H}+\epsilon\frac{a^{2}}{\nabla^{2}}\dot{\zeta})_{,i}-T_{\sigma}^{00}\frac{\dot{\zeta}}{H}].\label{eq:S_int_C}
\end{eqnarray}
The $\zeta$ cubic interaction and graviton actions can be found in
\cite{Maldacena:2002vr}.

Another commonly used gauge is the uniform curvature gauge, in which
\begin{equation}
h_{ij}=a^{2}(t)\left[e^{\gamma}\right]_{ij},\quad\gamma_{ii}=0,\quad\partial_{i}\gamma_{ij}=0.
\end{equation}
In this gauge, the inflaton degree of freedom is in $\delta\phi$.
However, this degree of freedom can be represented using the gauge-invariant
variable 
\begin{equation}
\zeta=-\frac{H}{\dot{\bar{\phi}}}\delta\phi^{(U)}\label{eq:zeta_U}
\end{equation}
 In this gauge, the ADM constraint renders 
\begin{equation}
N^{(1,U)}=-\epsilon\zeta,\quad N_{i}^{(1,U)}=\partial_{i}[\epsilon\frac{a^{2}}{\nabla^{2}}\dot{\zeta}]
\end{equation}
 We get the linear metric perturbation as 
\begin{equation}
\delta g_{\mu\nu}^{(U)}=\left(\begin{array}{cc}
2\epsilon\zeta & \epsilon\frac{a^{2}}{\nabla^{2}}\dot{\zeta}_{,i}\\
\epsilon\frac{a^{2}}{\nabla^{2}}\dot{\zeta}_{,i} & 0
\end{array}\right)\label{eq:del_g_U}
\end{equation}
 The free action is the same as in Eq.(\ref{eq:free-zeta-action}),
and $\sigma$-$\zeta$ cubic interaction action is 
\begin{eqnarray}
S_{\zeta\sigma\sigma}^{(U)} & = & \int d^{4}xa_{x}^{3}[T_{\sigma}^{00}\epsilon\zeta+T_{\sigma}^{0i}\epsilon\frac{a^{2}}{\nabla^{2}}\dot{\zeta}_{,i}].\label{eq:S_int_U}
\end{eqnarray}
From these perturbed actions, we can obtain the interaction Hamiltonian.
Particularly, note that up to the cubic interaction, $\mathcal{L}_{int}=-\mathcal{H}_{int}.$
Thus $S_{\zeta\sigma\sigma}=-\int dt\, H_{\zeta\sigma\sigma}(t).$

\section{Renormalization of Composite Operators\label{sec:Renormalization-of-Composite-Operators}}

In renormalized perturbation theory, one requires a regulator and
renormalization condition. In order to preserve the diffeomorphism
invariance, we need to adopt a covariant regulator. Here we choose
Pauli-Villars (PV) regulator, following \cite{Weinberg:2010wq,zinn}.
We will first review PV regularization in subsection \ref{sec:PV},
and renormalize $\sigma^{2}$ in subsection \ref{sec:ren-com-op}.
For correlators involving time integrals, we describes the adiabatic
expansion of time integral in subsection (\ref{sub:Adiabatic-Expansion}).

\subsection{Pauli-Villars Regularization}

\label{sec:PV}

We introduce a set of scalar regulator fields $\chi_{n}$ for $n=1,\cdots,s$
with the following free Lagrangian 
\begin{equation}
\mathcal{L}_{PV}=\sum_{n=1}^{s}C_{n}\left(-\frac{1}{2}g^{\mu\nu}\partial_{\nu}\chi_{n}\partial_{\nu}\chi_{n}-\frac{1}{2}M_{n}^{2}\chi_{n}^{2}\right).\label{eq:L-PV}
\end{equation}
The number of regulator fields $s$ depends on how many independent
divergences one need to remove. In order to eliminate UV divergences
up to some even order $2D,$we must take the $C_{n}$ and regulator
masses $M_{n}$ to satisfy 
\begin{eqnarray}
\sum_{N=0}^{s}C_{N}^{-1}=0 & ,\quad & \sum_{N=0}^{s}C_{N}^{-1}M_{N}^{2}=0,\cdots\label{eq:dummy25}\\
\sum_{n}^{s}C_{n}^{-1}M_{n}^{2D} & = & -m_{\sigma}^{2D}\label{eq:PV-reg}
\end{eqnarray}
where we used the notation $M_{0}^{2}=m_{\sigma}^{2}$ and $C_{0}=1$,
and let $\sigma_{0}=\sigma$ and $\sigma_{n}=\chi_{n}$. We use $\Lambda$
to represent the set of $M_{n}$, and the regulator dependence should
be removed by counter terms when $M_{n}$ goes to $\infty$ together. 

On a homogeneous FRW background, the physical and regulator scalar
field can be quantized as 
\begin{eqnarray}
[\sigma_{N},\dot{\sigma}_{M}] & = & ia^{-3}(t)\delta^{3}(\vec{x}-\vec{y})\delta_{NM}C_{N}^{-1}
\end{eqnarray}
 with the following mode decomposition 
\begin{eqnarray}
\sigma_{N}(\vec{x},t) & = & \int\frac{d^{3}k}{\left(2\pi\right)^{3}}(a_{N,\vec{k}}u_{N,\vec{k}}(t)+c.c)\\
{}[a_{N,\vec{p}},a_{M,\vec{k}}^{\dagger}] & = & \left(2\pi\right)^{3}C_{N}^{-1}\delta_{NM}\delta^{3}(\vec{k}-\vec{p}),
\end{eqnarray}
where $u_{N,\vec{p}}(t)$ satisfies the usual equation of motion 
\begin{equation}
\ddot{u}_{N,k}+3H\dot{u}_{N,k}+\left(\frac{k^{2}}{a^{2}}+M_{N}^{2}\right)u_{N,k}=0\label{eq:PV_EOM}
\end{equation}
with the Bunch-Davies initial condition
\begin{equation}
u_{N,k}(t)\to\frac{1}{\sqrt{2k}a(t)}\exp\left(-i\int^{t}\frac{k}{a(t')}dt'\right)\quad\mbox{for }t\to-\infty\label{eq:BD_cond}
\end{equation}
and Wronskian conditions%
\footnote{ Our treatment here differs from \cite{Weinberg:2010wq} in that the
physical scalar field $\phi$ here has no background solution, and
the regulator field $\chi_{n}$ does not mix with $\phi$ by mass
term. %
}
\begin{equation}
u_{N,k}\dot{u}_{N,k}^{*}-\dot{u}_{N,k}u_{N,k}^{*}=i/a^{3}.\label{eq:PV_Wronskian}
\end{equation}

Because $M_{n}\gg H$, Eq.(\ref{eq:PV_EOM}) possesses the WKB-type
solution
\begin{equation}
u_{n,k}(t)=\frac{1}{\sqrt{2\omega_{k}(t)a^{3}(t)}}\exp\left(-i\int^{t}\omega_{k}(t')dt'\right)\left[1+\frac{f_{1}(t)}{\omega_{k}(t)}+\frac{f_{2}(t)}{\omega_{k}^{2}(t)}+O(\omega_{k}^{-3})\right],
\end{equation}
where $\omega_{k}=\sqrt{k^{2}/a^{2}+M_{n}^{2}}$ and $f_{i}$ are
of zeroth order in $\omega_{k}$. Since we have to regulate up to
quadratic divergence in correlator computations, we need to know 
\begin{equation}
\left|u_{n,k}(t)\right|^{2}=\frac{1}{2\omega_{k}(t)a^{3}(t)}\left[1+\frac{2\mbox{Re}f_{1}(t)}{\omega_{k}(t)}+\frac{\left|f_{1}(t)\right|^{2}+2\mbox{Re}f_{2}(t)}{\omega_{k}^{2}(t)}+O(\omega_{k}^{-3})\right]\label{eq:PV_u2_form}
\end{equation}
up to second order. Due to the equation of motion (\ref{eq:PV_EOM}),
$f_{1}$ should satisfy
\begin{equation}
\frac{d}{dt}\left(\frac{f_{1}}{\omega_{k}}\right)=\frac{i}{2\omega_{k}}\left(\dot{H}+2H^{2}+\frac{1}{2}\frac{\left(\dot{H}+3H^{2}\right)M_{n}^{2}}{\omega_{k}^{2}}-\frac{5}{4}\frac{H^{2}M_{n}^{4}}{\omega_{k}^{4}}\right).
\end{equation}
Also, the Wronskian condition (\ref{eq:PV_Wronskian}) yields
\begin{equation}
\mbox{Re}f_{1}=0,\quad\left|f_{1}\right|^{2}+2\mbox{Re}f_{2}=\omega_{k}\frac{d}{dt}\left(\frac{\mbox{Im}f_{1}}{\omega_{k}}\right).
\end{equation}
 Then plugging these two results to Eq.(\ref{eq:PV_u2_form}) gives
\begin{equation}
\left|u_{n,k}\right|^{2}=\frac{1}{2\omega_{k}a^{3}}\left[1+\frac{\dot{H}+2H^{2}}{2\omega_{k}^{2}}+\frac{\left(\dot{H}+3H^{2}\right)M_{n}^{2}}{4\omega_{k}^{4}}-\frac{5H^{2}M_{n}^{4}}{8\omega_{k}^{6}}+O(\omega^{-3})\right].\label{eq:PV_u2}
\end{equation}

\subsection{Renormalization of Composite Operator}

\label{sec:ren-com-op}

The renormalization of composite operators in curved space-time is
the same as in flat space-time(see e.g. \cite{zinn,DeWitt:1975ys,birrell1982ix})
, just with new possible counter-terms made from curvature tensor.
For an operator of dimension $n$, one need to consider all possible
counter-terms of dimension $n$ or less. In our example model with
free massive scalar $\sigma$, we renormalize $\sigma^{2}$ as 
\begin{equation}
(\sigma^{2})_{r}=(\sigma+\sum_{n}\chi_{n})^{2}+\delta Z_{0}(\Lambda,m_{\sigma})+\delta Z_{1}(\Lambda,m_{\sigma})R,
\end{equation}
where $R$ is the Ricci scalar.

Next, we compute $\delta Z_{i}$'s divergent part. For example, let
us consider the one point function

\begin{equation}
\left\langle \left(\sigma^{2}\right)_{r}\right\rangle =\sum_{N=0}^{s}C_{N}^{-1}\int\frac{d^{3}k}{(2\pi)^{3}}\left|u_{N,k}\right|^{2}+\delta Z_{0}+\delta Z_{1}R.
\end{equation}
In order to determine the counter terms $\delta Z_{0}$ and $\delta Z_{1}$,
we introduce a comoving scale $Q$ such that $H\ll Q/a\ll M_{n}$
to break the Fourier space into the UV and the IR sector. Then we
use the WKB solution (\ref{eq:PV_u2}) for $k\gg Q$. Furthermore,
the contribution from the PV fields for $k\ll Q$ vanishes since it
is suppressed by $1/M_{n}$.

\begin{eqnarray}
\sum_{N}C_{N}^{-1}\int\frac{d^{3}k}{(2\pi)^{3}}\left|u_{N,k}\right|^{2} & = & \int^{Q}\frac{d^{3}k}{(2\pi)^{3}}\left|u_{0,k}\right|^{2}+\sum_{N=0}^{s}C_{N}^{-1}\int_{Q}^{\Lambda_{UV}}\frac{d^{3}k}{(2\pi)^{3}}\left|u_{i,k}\right|^{2}\nonumber \\
 & = & \int^{Q}\frac{d^{3}k}{(2\pi)^{3}}\left|u_{0,k}\right|^{2}+\frac{1}{48\pi^{2}}R\left(\ln\frac{a}{2Q}+\frac{10}{12}\right)\nonumber \\
 &  & -\frac{1}{96\pi^{2}}R\sum_{N=0}^{s}C_{N}^{-1}\ln M_{N}^{2}+\frac{1}{16\pi^{2}}\sum_{N=0}^{s}C_{N}^{-1}M_{N}^{2}\ln M_{N}^{2}.
\end{eqnarray}
Note that the arbitrary comoving scale $Q$ in the first two terms
should cancel each other.

In order to absorb the PV regulator dependence, we need
\begin{eqnarray}
\delta Z_{0} & = & \frac{1}{16\pi^{2}}\left[-\sum_{N}C_{N}^{-1}M_{N}^{2}\ln M_{N}^{2}+\mu_{0}^{2}\right],\\
\delta Z_{1} & = & \frac{1}{96\pi^{2}}\left[\sum_{N}C_{N}^{-1}\ln\frac{M_{N}^{2}}{\mu_{1}^{2}}\right],\label{eq:dZ1}
\end{eqnarray}
where $\mu_{0}$and $\mu_{1}$are unknown mass scales determined by
renormalization conditions. We set $\mu_{0}=0$ to have $\langle(\sigma)_{r}^{2}\rangle=0$
for flat space-time.

\subsection{Adiabatic Expansion of Time Integral\label{sub:Adiabatic-Expansion}}

In order to compute some correlators using the in-in formalism (\ref{eq:in-in_formalism}),
such as two-point function $\left\langle \sigma^{2}\zeta\right\rangle $,
we need to integrate PV field contributions over time. In this subsection,
we present how to calculate the time integral of PV fields by adiabatically
expanding the integral. 

For simplicity, consider a diagram with one internal vertex. Using
the WKB solution (\ref{eq:PV_u2}) of a PV field, the general form
of the time integral is 
\begin{equation}
I(k_{1},k_{2},\cdots,t_{f})=\int_{-\infty}^{t_{f}}dt\, G(k_{1},k_{2},\cdots;t_{f},t)e^{-i\int_{t}^{t_{f}}\omega(t')dt'},
\end{equation}
where $\omega(t)=\omega_{k_{1}}(t)+\omega_{k_{2}}(t)+\cdots$ and
$G(k_{1},k_{2},\cdots;t_{f},t)=O(\omega^{n})$. Because the integrand
is a rapidly oscillatory function, the dominant contribution comes
near the final time $t_{f}$. Thus, using integration by parts we
expand the integral with respect to $\omega$:
\begin{eqnarray}
I(k_{1},k_{2},\cdots,t_{f}) & = & \frac{G(k_{1},k_{2},\cdots;t_{f},t_{f})}{i\omega(t_{f})}-\int_{-\infty}^{t_{f}}dt\,\left(\frac{d}{dt}\frac{G(k_{1},k_{2},\cdots;t_{f},t)}{i\omega(t)}\right)e^{-i\int_{t}^{t_{f}}\omega(t')dt'}\label{eq:dummy27}\\
 & = & \frac{G(k_{1},k_{2},\cdots;t_{f},t_{f})}{i\omega(t_{f})}-\left.\left(\frac{1}{i\omega(t)}\frac{d}{dt}\frac{G(k_{1},k_{2},\cdots;t_{f},t)}{i\omega(t)}\right)\right|_{t=t_{f}}\nonumber \\
 &  & +\left.\left[\frac{1}{i\omega(t)}\frac{d}{dt}\left(\frac{1}{i\omega(t)}\frac{d}{dt}\frac{G(k_{1},k_{2},\cdots;t_{f},t)}{i\omega(t)}\right)\right]\right|_{t=t_{f}}+O(\omega^{n-4}).\label{eq:adiabatic_exp}
\end{eqnarray}
Note that the mode functions $u_{n,k}$ and $u_{n,k}^{*}$ appear
in pairs because of Wick contraction. Hence, the final result should
be written in terms of $\left|u_{n,k}(t_{f})\right|^{2}$ and their
time derivatives, and we can compute the time integral up to arbitrary
order of $\omega$. It is straightforward to generalize this to the
cases with any number of internal vertices.

\section{Two-Point Function $\left\langle \left(\sigma^{2}\right)_{r}\zeta\right\rangle $
in the Uniform Curvature Gauge\label{sec:Two-Point-Functions-in-UG}}

In this section, we compute $\left\langle \left(\sigma^{2}\right)_{r}\zeta\right\rangle $
using the uniform curvature gauge in the quasi-de Sitter(dS) background,
where the slow-roll factor $\epsilon$ is constant. Then we will show
that the results in the both gauges are consistent with each other.
Particularly, for the massless limit, the next leading order term
in the uniform curvature gauge that indeed decays as $p^{2}/a^{2}$. 

The two-point function is the same as in the comoving gauge except
that the counter term contribution appears in the leading order. 
\begin{eqnarray}
\widetilde{\left\langle \left(\sigma^{2}\right)_{r}\zeta\right\rangle _{p}^{U}} & = & \int d^{3}x\, e^{-i\vec{p}\cdot\vec{x}}\int^{t}d^{4}z\, a^{3}(t_{z})\sum_{N=0}^{n}\left\langle \left[\sigma_{N}^{2}(t,\vec{x})\zeta(t,\vec{0}),\frac{i}{2}\left(T_{\sigma}^{\mu\nu}\delta g_{\mu\nu}^{(U)}\right)_{z}\right]\right\rangle \nonumber \\
 &  & +\delta Z_{1}\widetilde{\left\langle R\zeta\right\rangle _{p}},\label{eq:sigma2zeta_UG-org}
\end{eqnarray}
where $R$ is the Ricci scalar. After taking non-derivate interaction
term $T_{\sigma}^{00}\delta g_{00}^{(U)}$ only, factoring $\epsilon$
and $\zeta$ out from the integral, we get 
\begin{eqnarray}
\widetilde{\left\langle \left(\sigma^{2}\right)_{r}\zeta\right\rangle _{p}^{U}} & = & i\left|\zeta_{p}^{o}\right|^{2}\epsilon\int^{t}d^{4}z\, a^{3}(t_{z})\sum_{N=0}^{n}\left\langle \left[\sigma_{N}^{2}(t,\vec{x}),\left(T_{\sigma}^{00}\right)_{z}\right]\right\rangle \nonumber \\
 &  & +24\epsilon H^{2}\left|\zeta_{p}^{o}\right|^{2}\delta Z_{1}+O\left(\dot{\epsilon},\epsilon^{2},\frac{p^{2}}{a^{2}}\right),
\end{eqnarray}
where we have used the perturbed curvature in the uniform curvature
gauge 
\begin{equation}
R=12H^{2}-6\epsilon H^{2}+24\epsilon H^{2}\zeta+4\epsilon H\dot{\zeta}+\cdots,
\end{equation}
where $\cdots$ denotes $O(\dot{\epsilon},\epsilon^{2})$ terms or
terms proportional to the equation of motion of $\zeta$.

Since $T_{\sigma}^{00}=\mathcal{L}_{\sigma}+\sum_{N}\left[\left(\frac{\nabla}{a}\sigma_{N}\right)^{2}+M_{N}^{2}\sigma_{N}^{2}\right]$,
together with the identities (\ref{eq:Res_Int_2}),(\ref{eq:Res_Int_1}),
and 
\begin{equation}
i\int^{t}d^{4}z\, a^{3}(t_{z})\left\langle \left[\sigma_{N}^{2}(t,\vec{x}),\sigma_{N}^{2}(z)\right]\right\rangle =-2\frac{\partial}{\partial M_{N}^{2}}\left\langle \left(\sigma_{N}^{2}\right)_{p}\right\rangle ,
\end{equation}
with $T_{\sigma}^{00}=\mathcal{L}_{\sigma}+\sum_{N}\left[\left(\frac{\nabla}{a}\sigma_{N}\right)^{2}+M_{N}^{2}\sigma_{N}^{2}\right]$,
we have
\begin{eqnarray}
\widetilde{\left\langle \left(\sigma^{2}\right)_{R}\zeta\right\rangle _{p}^{U}}+\frac{1}{H}\frac{d}{dt}\left\langle \left(\sigma^{2}\right)_{R}\right\rangle \widetilde{\left\langle \zeta\zeta\right\rangle _{p}} & = & \sum_{N}F_{N}(t)+O\left(\dot{\epsilon},\epsilon^{2},\frac{p^{2}}{a^{2}}\right),\label{eq:sigma2zeta_UG}\\
F_{N}(t) & = & \epsilon\left(2\left\langle \left(\sigma_{N}^{2}\right)_{p}\right\rangle -Z_{N}^{-1}\left.\frac{k^{3}}{2\pi^{2}}\left|u_{N,k}\right|^{2}\right|_{p}^{\Lambda_{UV}}-2M_{N}^{2}\frac{\partial}{\partial M_{N}^{2}}\left\langle \left(\sigma_{N}^{2}\right)_{p}\right\rangle \right)\nonumber \\
 &  & +\frac{1}{H}\frac{d}{dt}\left\langle \sigma_{N}^{2}\right\rangle .
\end{eqnarray}
Although the rhs of Eq.(\ref{eq:sigma2zeta_UG-org}) is well-defined
and regulator independent, individual terms are not. Thus, we insert
counter terms to have each term regulator independent
\begin{eqnarray}
\sum_{N}F_{N}(t) & = & \epsilon\left(2\left\langle \left(\sigma^{2}(t)\right)_{r,p}\right\rangle +\frac{p^{3}}{2\pi^{2}}\left|u_{p}(t)\right|^{2}-2m_{\sigma}^{2}\frac{\partial}{\partial m_{\sigma}^{2}}\left\langle \left(\sigma^{2}(t)\right)_{r,p}\right\rangle \right)\nonumber \\
 &  & +\frac{1}{H}\frac{d}{dt}\left\langle \left(\sigma^{2}\right)_{r}\right\rangle ,\label{eq:FN_UG}
\end{eqnarray}
where we have put the counter terms $\delta Z_{0}$ and $\delta Z_{1}R$
into each one-point function, and the PV field contribution from the
third term cancels with those from the other terms. Then, using the
relation (\ref{eq:Gauge_Inv}) one can find the rhs of Eq.(\ref{eq:FN_UG})
is consistent with the result (\ref{eq:sigma2zeta_ds_CG}) in the
comoving gauge in the quasi-dS background after explicitly computing
renormalized one-point function $\left\langle \left(\sigma^{2}(t)\right)_{r,p}\right\rangle $.
On the other hand, the rhs does not depend on the renormalization
as all counter terms cancel. Hence, we can arrive at the same conclusion
using the one point function using super-horizon approximation in
the dS space-time,
\begin{equation}
\left\langle \left(\sigma^{2}(t)\right)_{r,p}\right\rangle \approx\int_{p}^{caH}\frac{d^{3}k}{\left(2\pi\right)^{3}}\left|u_{k}(t)\right|^{2}\approx\int_{p}^{caH}\frac{d^{3}k}{\left(2\pi\right)^{3}}\frac{\left|\Gamma(\nu)\right|^{2}}{4\pi Ha^{3}}\left(\frac{k}{2aH}\right)^{-2\nu},
\end{equation}
where the arbitrary constant $c\lesssim O(1)$. Note that the UV boundary
of the integral should be a comoving scale in order to to keep the
spatial dilatation symmetry.

\subsection*{Massless Limit}

For the massless limit $m_{\sigma}^{2}/H^{2}\ll\ln p/aH$, we can
compute the two-point function explicitly without neglecting any gravitational
couplings. We calculate up to the next leading term here. We decompose
Eq. (\ref{eq:sigma2zeta_UG-org}) as
\begin{equation}
\widetilde{\left\langle \left(\sigma^{2}\right)_{r}\zeta\right\rangle _{p}^{U}}=I_{0}(p,t)+\sum_{n=1}^{s}I_{n}(p,t)+I_{c.t.}(p,t),\label{eq:sigma2zeta-decomposed}
\end{equation}
where $I_{0},\, I_{n},$ and $I_{c.t.}$ are the contributions from
the physical field $\sigma$, the PV field $\chi_{i}$ and the counter
terms, respectively. Since all the gravitational couplings are $O(\epsilon)$
(See Eq. (\ref{eq:S_int_U})), we may use the mode functions $\zeta_{p}$
and $u_{k}$ in the pure dS for $O(\epsilon)$ correction to the two-point
function. Then a long but straightforward calculation gives
\begin{eqnarray}
I_{0}(p,t) & = & \int d^{3}x\, e^{-i\vec{p}\cdot\vec{x}}\int^{t}d^{4}z\, a^{3}(t_{z})\sum_{N=0}^{n}\left\langle \left[\sigma^{2}(t,\vec{x})\zeta(t,\vec{0}),\frac{i}{2}\left(T_{\sigma}^{\mu\nu}\delta g_{\mu\nu}^{(U)}\right)_{z}\right]\right\rangle \label{eq:dummy28}\\
 & = & \frac{1}{4\pi^{2}}\epsilon H^{2}\left|\zeta_{p}^{o}\right|^{2}\left[-\frac{1}{3}\frac{p^{3}}{a^{3}H^{3}}\frac{\Lambda}{aH}+2\log\frac{\Lambda}{p}+\frac{5}{3}\frac{p^{2}}{a^{2}H^{2}}\log\frac{\Lambda}{p}+1-\frac{p^{2}}{a^{2}H^{2}}+O\left(\frac{p^{4}}{a^{4}H^{4}}\right)\right]\label{eq:I0_UG}\\
 &  & +O(\epsilon^{2},\dot{\epsilon}).\nonumber 
\end{eqnarray}
The PV field contribution $I_{n}$ requires some more technical explanation.
If we write the WKB solution (\ref{eq:PV_u2_form}) as 
\begin{equation}
u_{n,k}(t)=\alpha_{k}(t)e^{-i\int^{t}w_{k}(t')dt'},\label{eq:PV_u2_alpha}
\end{equation}
 the PV field contribution $I_{n}$ is written as
\begin{eqnarray}
I_{n}(p,t) & = & C_{n}^{-1}\int_{Q}\frac{d^{3}k_{1}}{(2\pi)^{3}}d^{3}k_{2}\delta^{(3)}(\vec{k}_{1}+\vec{k}_{2}-\vec{p})\,\mbox{Im}\left[\int^{t}dt_{z}e^{i\int_{t}^{t_{z}}\left(\omega_{k_{1}}(t')+\omega_{k_{2}}(t')\right)dt'}G_{n}(k_{1},k_{2};t,t_{z})\right],\label{eq:In}
\end{eqnarray}
where
\begin{eqnarray}
G_{n}(k_{1},k_{2};t,t_{z}) & = & -2a_{z}^{3}\zeta_{p}(t)\alpha_{k_{1}}(t)\alpha_{k_{2}}(t)\left(\sum_{i}\widehat{O_{i}}\right)\zeta_{p}^{*}(t_{z})\alpha_{k_{1}}^{*}(t_{z})\alpha_{k_{2}}^{*}(t_{z}),\\
\left(\widehat{O_{1}}\right) & = & \frac{1}{2}\left[\left(i\omega_{k_{1}}(t_{z})+\partial_{t_{z}}^{(1)}\right)\left(i\omega_{k_{2}}(t_{z})+\partial_{t_{z}}^{(2)}\right)-\frac{\vec{k}_{1}\cdot\vec{k}_{2}}{a_{z}^{2}}+M_{n}^{2}\right]\left(2\epsilon\right),\\
\left(\widehat{O_{2}}\right) & = & \left[\frac{\vec{k}_{2}\cdot\vec{p}}{a_{z}^{2}}\left(i\omega_{k_{1}}(t_{z})+\partial_{t_{z}}^{(1)}\right)+\frac{\vec{k}_{3}\cdot\vec{p}}{a_{z}^{2}}\left(i\omega_{k_{2}}(t_{z})+\partial_{t_{z}}^{(2)}\right)\right]\left(\epsilon\frac{a_{z}^{2}}{p^{2}}\partial_{t_{z}}^{\zeta}\right),
\end{eqnarray}
where $\partial_{t_{z}}^{(i)}$ and $\partial_{t_{z}}^{\zeta}$ denotes
the time derivative with respect to $\alpha_{k_{i}}^{*}(t_{z})$ and
$\zeta_{p}^{*}(t_{z})$, respectively, and $\left(\widehat{O_{1}}\right)$
and $\left(\widehat{O_{1}}\right)$ correspond to the (00) and the
(i0) components of the gravitational couplings, respectively. Notice
that $\alpha_{k}=O\left(\omega^{-1/2}\right)$ and $G(k_{1},k_{2};t,t_{z})=O(\omega^{0})$,
and thus $I_{n}$ has quadratic divergences superficially. However,
the quadratic divergences arising from $\left(\widehat{O_{1}}\right)$
vanish in the $M_{n}\to\infty$ limit. Effectively, the integral (\ref{eq:In})
is linearly divergent. That means we have to adiabatically expand
the integral to the second order. Similarly, the integral of the two-point
function in the comoving gauge is quadratic divergent, and thus one
need to expand the integral to the third order. This makes the computation
easier in the uniform curvature gauge. Using
\begin{eqnarray}
\left|\alpha_{k}(t)\right|^{2} & = & \frac{1}{2\omega_{k}a^{3}}\left[1+\beta_{2}(k,t)+O(\omega_{k}^{-3})\right],\\
\alpha_{k}(t)\dot{\alpha}_{k}^{*}(t) & = & \frac{1}{2\omega_{k}a^{3}}\left[\gamma_{0}(k,t)-i\omega_{k}\beta_{2}(k,t)+O(\omega_{k}^{-2})\right],\\
\alpha_{k}(t)\ddot{\alpha}_{k}^{*}(t) & = & \frac{1}{2\omega_{k}a^{3}}\left[-3iH-2i\gamma_{0}(k,t)+i\frac{k^{2}/a^{2}}{\omega_{k}^{2}}H+O(\omega_{k}^{-1})\right],
\end{eqnarray}
where
\begin{eqnarray}
\gamma_{0}(k,t) & = & -\frac{3}{2}H+\frac{1}{2}\frac{k^{2}/a^{2}}{\omega_{k}^{2}}H,\\
\beta_{2}(k,t) & = & \frac{\dot{H}+2H^{2}}{2\omega_{k}^{2}}+\frac{\left(\dot{H}+3H^{2}\right)M_{n}^{2}}{4\omega_{k}^{4}}-\frac{5H^{2}M_{n}^{4}}{8\omega_{k}^{4}},
\end{eqnarray}
which are obtained by combining Eq. (\ref{eq:PV_u2}) with Eq. (\ref{eq:PV_u2_alpha}),
the integral (\ref{eq:In}) becomes
\begin{eqnarray}
I_{n}(p,t) & = & \frac{C_{n}^{-1}}{4\pi^{2}}\epsilon H^{2}\left|\zeta_{p}^{o}\right|^{2}\left[-\frac{1}{3}\frac{p^{3}}{a^{3}H^{3}}\frac{\Lambda}{aH}+2\log\frac{2\Lambda}{aM_{n}}+\frac{5}{3}\frac{p^{2}}{a^{2}H^{2}}\log\frac{2\Lambda}{aM_{n}}-\frac{5}{3}\right.\nonumber \\
 &  & \qquad\qquad\quad\left.-\frac{25}{18}\frac{p^{2}}{a^{2}H^{2}}+O\left(\frac{p^{4}}{a^{4}H^{4}}\right)\right]+O(\epsilon^{2},\dot{\epsilon}).\label{eq:In_UG}
\end{eqnarray}
Note that all $\Lambda$ dependent terms in $I_{0}+\sum_{n}I_{n}$
vanishes by the PV field normalization conditions (\ref{eq:PV-reg}). 

Putting Eqs. (\ref{eq:I0_UG}) and (\ref{eq:In_UG}) together into
Eq. (\ref{eq:sigma2zeta-decomposed}), we have
\begin{eqnarray}
\widetilde{\left\langle \left(\sigma^{2}\right)_{R}\zeta\right\rangle _{p}^{U}} & = & \frac{1}{4\pi^{2}}\epsilon H^{2}\left|\zeta_{p}^{o}\right|^{2}\left[2\log\frac{a\mu_{1}}{2p}+\frac{5}{3}\frac{p^{2}}{a^{2}H^{2}}\log\frac{a\mu_{1}}{2p}+\frac{8}{3}\right.\nonumber \\
 &  & \qquad\qquad\qquad\left.+\frac{7}{18}\frac{p^{2}}{a^{2}H^{2}}+O\left(\frac{p^{4}}{a^{4}H^{4}}\right)\right]+O(\epsilon^{2},\dot{\epsilon}).
\end{eqnarray}
We still need to compute one-point function $\frac{d}{dt}\left\langle \left(\sigma^{2}\right)_{r}\right\rangle $
up to $O(\epsilon)$ in order to compare the results in both gauges.
Because mode functions for a massless scalar field are $O(\epsilon^{0})$,
we need $O(\epsilon)$ correction on it. In a quasi-dS background,
we take an ansatz for the mode function
\begin{equation}
u_{k}(t)=\left(\frac{1}{\sqrt{2k}a(t)}+i\frac{H(t)}{\sqrt{2k^{3}}}\right)e^{i\frac{k}{a(t)H(t)}}+\frac{\epsilon(t)}{\sqrt{2k}a(t)}f_{k}(t)e^{i\frac{k}{a(t)H(t)}},
\end{equation}
where $f_{k}(t)=O(\epsilon^{0})$ so that it recovers the dS solution
in the $\epsilon\to0$ limit. Applying this to the differential equation
\begin{equation}
\ddot{u}_{k}(t)+3H\dot{u}_{k}(t)+\frac{k^{2}}{a^{2}}u_{k}(t)=0,
\end{equation}
we get
\begin{equation}
\ddot{f_{k}}+\left(H(t)-2i\frac{k}{a(t)}\right)\dot{f}_{k}-H(t)^{2}f_{k}=3H(t)^{2}-2i\frac{k}{a(t)}H(t)-2\frac{k^{2}}{a(t)^{2}}+O(\epsilon),
\end{equation}
whose solution is 
\begin{eqnarray}
f_{k}\left(t\right) & = & -\frac{3}{2}+iq+\frac{i}{2}\frac{1}{q}+\left(1-\frac{i}{q}\right)e^{-2iq}Ei(2iq)\\
 &  & +c_{1}\left(1+\frac{i}{q}\right)+c_{2}\left(1-\frac{i}{q}\right)e^{-2iq}
\end{eqnarray}
where $q=\frac{k}{a(t)H(t)}$, and $Ei$ is the exponential integral
function
\begin{eqnarray}
Ei(z) & = & -\int_{-z}^{\infty}\frac{e^{-t}}{t}dt\\
Ei(\pm ix\to\infty) & \to & \pm i\pi+e^{\pm ix}\left(\frac{0!}{(\pm ix)}+\frac{1!}{(\pm ix)^{2}}+\frac{2!}{(\pm ix)^{3}}+\cdots\right).
\end{eqnarray}
Matching this solution with the Bunch-Davies initial condition (\ref{eq:BD_cond})
and the Wronskian condition (\ref{eq:PV_Wronskian}) respectively
give
\begin{equation}
c_{2}=-i\pi\:\mbox{and \,\ }c_{1}=\frac{1}{2}.\label{eq:dummy29}
\end{equation}
Then the mode function with $O(\epsilon)$ correction in a quasi-dS
space-time becomes
\begin{eqnarray}
u_{k}(t) & = & \left(\frac{1}{\sqrt{2k}a}+i\frac{H}{\sqrt{2k^{3}}}\right)e^{i\frac{k}{aH}}\\
 &  & +\frac{\epsilon}{\sqrt{2k}a}\left[-1+i\frac{k}{aH}+i\frac{aH}{k}+\left(1-i\frac{aH}{k}\right)\left(-i\pi+Ei(2i\frac{k}{aH})\right)e^{-2i\frac{k}{aH}}\right]e^{i\frac{k}{aH}}+O(\epsilon^{2},\dot{\epsilon}).\label{eq:uk}
\end{eqnarray}
Now we calculate the one-point function using this mode function as
shown in Subsection \ref{sec:ren-com-op}, and we get
\begin{equation}
\frac{d}{dt}\left\langle \left(\sigma^{2}\right)_{r}\right\rangle =\frac{H^{3}}{4\pi^{2}}+\frac{\epsilon H^{3}}{2\pi^{2}}\left(\log\frac{H}{\mu_{1}}+\frac{1}{6}-\gamma\right)+O(\epsilon^{2},\dot{\epsilon}).\label{eq:Renormed_1pt}
\end{equation}

Finally, we find
\begin{eqnarray}
\frac{1}{H}\frac{d}{dt}\left\langle \left(\sigma^{2}\right)_{r}\right\rangle \widetilde{\left\langle \zeta\zeta\right\rangle _{p}}+\widetilde{\left\langle \left(\sigma^{2}\right)_{r}\zeta\right\rangle _{p}^{U}} & = & \frac{H^{2}(t)}{4\pi^{2}}\left|\zeta_{p}(t)\right|^{2}+\frac{\epsilon H^{2}}{2\pi^{2}}\left|\zeta_{p}^{o}\right|^{2}\left[\log\frac{aH}{2p}+\frac{3}{2}-\gamma\right]\nonumber \\
 &  & +\frac{\epsilon H^{2}}{4\pi^{2}}\left|\zeta_{p}^{o}\right|^{2}\frac{p^{2}}{a^{2}H^{2}}\left[\frac{13}{18}-2\gamma+\frac{5}{3}\log\frac{a\mu_{1}}{2p}+2\log\frac{H}{\mu_{1}}\right]\nonumber \\
 &  & +O\left(\epsilon^{2},\dot{\epsilon},\frac{p^{4}}{a^{4}H^{4}}\right).
\end{eqnarray}
The non-$p^{2}/a^{2}$-suppressed terms are rewritten as
\begin{eqnarray}
\frac{H^{2}(t)}{4\pi^{2}}\left|\zeta_{p}^{o}\right|^{2}+\frac{\epsilon H^{2}}{2\pi^{2}}\left|\zeta_{p}^{o}\right|^{2}\left[\log\frac{aH}{2p}+\frac{3}{2}-\gamma\right] & \approx & \frac{H^{2}(t)}{4\pi^{2}}\left|\zeta_{p}^{o}\right|^{2}\times\left(1+2\epsilon\log\frac{aH}{p}\right)\\
 & \approx & \frac{H^{2}(t)}{4\pi^{2}}\left|\zeta_{p}^{o}\right|^{2}\left(\frac{p}{aH}\right)^{-2\epsilon}\\
 & \approx & \frac{H_{*}^{2}}{4\pi^{2}}\left|\zeta_{p}^{o}\right|^{2}.
\end{eqnarray}
As expected, this is the result (\ref{eq:sigma2zeta_ds_CG}) in the
comoving gauge. The other terms are suppressed by $ $the factor $p^{2}/a^{2}$.
This explicitly proves that the next leading terms for the two-point
function $\left\langle \left(\sigma^{2}\right)_{r}\zeta\right\rangle $
are $O(p^{2}/a^{2})$.

\bibliographystyle{kp}
\bibliography{WardId-citations,ConsistencyRelation,Curvaton,Composite_operator,Non-Gaussianity,misc}

\end{document}